\documentclass[english,pra,showpacs,showkeys,tightenlines,secnumarabic,11pt]{revtex4}
\usepackage[T1]{fontenc}
\usepackage[latin1]{inputenc}
\usepackage{amsmath}
\usepackage{graphicx}
\usepackage{amssymb}
\usepackage{epsfig}
\usepackage{graphics}
\usepackage[mathscr]{euscript}
\usepackage{psfrag}
\usepackage{pstricks}
\usepackage{pst-node}
\usepackage{babel}

\begin{document}
\title{Collisions of protons with light nuclei shed new light on nucleon structure}
\author{D. Treleani}
\email{daniele.treleani@ts.infn.it} \affiliation{ Dipartimento di
Fisica del{{l'}}Universit\`a di Trieste and INFN, Sezione di
Trieste,\\ Strada Costiera 11, Miramare-Grignano, I-34151 Trieste,
Italy.}
\author{G. Calucci}
\email{giorgio.calucci@ts.infn.it} \affiliation{ Guest at the Dipartimento di Fisica del{{l'}}Universit\`a, Trieste}

\begin{abstract}
The high rates of multi-parton interactions at the LHC can provide a unique opportunity to study the multi-parton structure of the hadron. To this purpose high energy collisions of protons with nuclei are particularly suitable. The rates of multi-parton interactions depend in fact both on the partonic multiplicities and on the distributions of partons in transverse space, which produce different  effects on the cross section in $pA$ collisions, as a function of the atomic mass number $A$. Differently with respect to the case of multi-parton interactions in $pp$ collisions, the possibility of changing the atomic mass number provides thus an additional handle to distinguish the diverse contributions.
Some relevant features of double parton interactions in $pD$ collisions have been discussed in a previous paper. In the present paper we show how the effects of double and triple correlation terms of the multi-parton structure can be disentangled, by comparing the rates of multiple parton interactions in collisions of protons with $D,\ ^3H$ and $^3He$.
\end{abstract}


\maketitle

\section{Introduction}

The experimental evidence\cite{Akesson:1986iv}\cite{Abe:1997bp}\cite{Abe:1997xk}\cite{Abazov:2009gc}\cite{Abazov:2011rd}\cite{ATLAS} and the beginning of the operations at the Large Hadron Collider (LHC) have recently triggered a lot of attention to the problem of Multiple Parton Interactions (MPI) in high energy $pp$ collisions. Several papers have been written on the topic in the last few months and four international workshops have been organized on the theme\cite{DESY2011}\cite{Glasgow}\cite{Bartalini:2011jp}\cite{DESY}\cite{Bartalini:2010su}\cite{Perugia}. Issues discussed in the literature range from estimates of the contributions of MPI in various reaction channels of particular interest for the LHC physics\cite{Maina:2009sj}\cite{Maina:2009vx}\cite{Gaunt:2010pi}\cite{Hussein:2010zz}\cite{Bandurin:2010gn}\cite{Maina:2010vh}\cite{Baranov:2011ch}\cite{Berger:2011ep}\cite{Novoselov:2011ff}\cite{Quackenbush:2011bf}\cite{Myska:2011ji}\cite{Kom:2011nu} to the effects on the global features of the inelastic event and of the underlying event\cite{Corke:2009tk}\cite{d'Enterria:2010hd}\cite{Berger:2009cm}\cite{Field:2010bc}\cite{Bartalini:2011xj}\cite{Deak:2010gk}\cite{Deak:2011ga}\cite{Bartels:2011qi}\cite{Grothe:2011ty}, the QCD evolution of the double parton distributions\cite{Snigirev:2003cq}\cite{Gaunt:2009re}\cite{Snigirev:2010ds} and the general formulation of MPI within QCD\cite{Diehl:2010dr}\cite{Blok:2010ge}\cite{Diehl:2011tt}\cite{Diehl:2011yj}\cite{Diehl:2011fp}\cite{Gaunt:2011xu}\cite{Gaunt:2011xd}\cite{Blok:2011bu}\cite{Ryskin:2011kk}\cite{Dokshitzer:2012it}\cite{Manohar:2012jr}\cite{Manohar:2012pe}\cite{Gaunt:2012wv}\cite{Ryskin:2012qx}. Somewhat less attention has been devoted to the study of MPI in hadron-nucleus collisions, although all effects of MPI are sizably enhanced in that case as a consequence of the much larger parton flux\cite{Frankfurt:2004kn}\cite{Cattaruzza:2004qb}\cite{Strikman:2010bg}. In our opinion, a good reason to pay more attention to $pA$ collisions in this context is that, when studied jointly with $pp$, MPI in $pA$ collisions can provide a unique handle to study some aspects of the multi-parton structure of the hadron\cite{Strikman:2001gz}.

In spite of being directly related to the multi-parton distribution functions, MPI in $pp$ collisions can in fact provide only a partial information on the multi-parton distributions. Because of the localization of the large momentum transfer processes, the incoming parton flux, and thus the  multi-parton distribution functions, depend explicitly on the relative transverse distances between the interacting parton pairs\cite{Paver:1982yp}\cite{Paver:1984ux}. At the same time, correlations in the hadron structure will prevent expressing the multi-parton distributions as an uncorrelated product of one-body distribution functions\cite{Calucci:1997ii}. The rates of MPI will therefore depend both on the typical relative transverse parton distances and also on the moments of the multi-parton distribution in multiplicity. In the case of MPI in $pp$ collisions, the two features are unavoidably linked in the measured cross section\cite{Strikman:2001gz}\cite{Calucci:2010wg} and, as a consequence, only a partial information on the multi-parton distributions can be obtained by measuring MPI in $pp$ collisions.

On the other hand, MPI in $pA$ collisions can provide a further handle for a deeper insight into the correlated multi-parton structure\cite{Strikman:2001gz}\cite{Calucci:2010wg}.
    In $pA$ collisions the MPI cross section is a function of the multiplicity of the target nucleons.
    In the case of two or more target nucleons, the dimensional scale factor, characterizing each MPI event, is provided both by the hadronic parameters, radius and partonic correlation length, and by the nuclear size. The hadronic scale measured with the generalized parton distributions\cite{Frankfurt:2003td} is rather small as compared to the nuclear scale, which thus acquires a dominant role even in the case of light nuclei. When two or more target nucleons take part to the hard interaction, the contributions to the double parton scattering cross section in $pA$ collisions depend therefore only weakly on the hadronic dimensions and, by studying MPI in $pA$ collisions, one can thus single out the effects of the moments in multiplicity of the multi-parton distributions from the effects due to the correlations in the transverse parton coordinates.

Studying MPI in $pA$ is simpler in the case of light nuclei, where the binding is not very strong: in this case the structure of the nucleon is not much affected by the binding and the non-relativistic form of the wave function in the rest frame of the nucleus is appropriate. In such a case it's simple enough to construct the boost, which allows to move from the rest frame of the nucleus, where the wave function is given, to the hadron-nucleon rest frame, which is most suitable to describe the collision and where the relativistic expression of the nuclear wave function is compulsory\cite{Karmanov}\cite{Calucci:2010wg}. Some relevant features of double parton collisions in proton-Deuteron interactions have been discussed in a previous article\cite{Calucci:2010wg}. In the present paper we will extend the analysis to the case of double and triple parton collisions of protons with Deuterium, $^3H$ and $^3He$; the main results presented in \cite{Calucci:2010wg} will be reviewed and completed.
 \par
The paper is organized as follows: the processes are described in a covariant way, using the formalism of Feynman graphs supplemented by effective vertices for the non perturbative dynamics. The resulting expression is then reduced to a form containing the fractional longitudinal momenta and the transverse coordinates. The use of the non-relativistic nuclear wave function in the relativistic process is justified by the same argument of our previous study of double parton interactions in $pD$ collisions\cite{Calucci:2010wg}. The only difference concerns the technicalities, which are heavier when considering the three-body dynamics of a non-relativistic nuclear bound state (the problem is discussion in detail in Appendix A). Concerning the double and triple parton distributions, to remain as general as possible, we have not introduced any explicit expression with correlation parameters. Rather we have limited our discussion to the actual relations between the observables (namely the MPI cross sections) and non perturbative quantities, characterizing the double and triple parton distributions, directly related to the correlated multi-parton structure: Namely the various overlap integrals, with a strong dependence on the partonic correlations in transverse space, and the two functions of fractional momenta (one for the double and one for the triple multi-parton distributions) representing the deviation of the parton population from an uncorrelated, i.e. Poissonian, distribution. A simple and fully explicit correlated Gaussian model, where all quantities are worked out in detail, is presented in Appendix B.

\par We like to anticipate a feature, which we will meet more than once and has no analogy in the case of MPI in $pp$ collisions. The spread of the momenta of the bound nucleons will allow to produce the same initial partonic configuration in different ways. MPI in $pA$ collisions is thus characterized by quantum interferences between initial state configurations, which differ in the nuclear fractional momenta and in the transverse parton coordinates.

\par
Since our main interest is to recognize the most important features of MPI in collisions of protons with light nuclei, we introduce a drastic simplification in treating all the particles entering the game as spinless bosons. In addition other finer details are neglected from the beginning: the proton and neutron masses are considered equal, as well as the binding energies of $^3H$ and $^3He$, so $p\ ^3H$  and $p\ ^3He$  collisions will be considered as equal.\par
In Section 2 the double parton cross sections are worked out, distinguishing the cases with a different number of spectator nucleons. In Section 3 the triple parton cross sections are worked out, distinguishing the cases with the same pocedure. In Section 4 we discuss the relations between the accessible experimental information, namely the different cross sections, and the unknown quantities most directly related to the multi-parton correlations. The main points examined and the results obtained are finally summarized in the last part of the paper.

   \section {Double scattering on deuteron or Tritium}
  \subsection {Only one bound nucleon interacts with large momentum transfer}

Some aspects of double parton scattering of protons with Deuterium have been already discussed in\cite{Calucci:2010wg}. The process will be reviewed here and the analysis will be extend to the case of double parton collisions of protons with $^3H$ and $^3He$. In a double parton scattering on a Deuteron one has two possibilities, either only one nucleon interacts with large transverse momentum exchange or the interacting nucleons are two. Analogously, in Tritium one may have either one or two spectator nucleons. With minor adjustments, the case of Tritium can thus be reduced to the case of Deuteron.\par
The analytical expression for the hard scattering, when one of the component nucleons interacts twice and there are one (Deuteron) or two (Tritium) spectators, is conveniently expressed through the discontinuity of the forward scattering amplitude (see Fig.1). We start with the discontinuity of the amplitude ${\cal F}_2$ of double scattering between two free nucleons:

\begin{eqnarray}
{\rm Disc}\;{\cal F}_2 & (L,&L',D,D')=\frac{1}{(2\pi)^{18}}\int\frac{{\hat\phi}_p}{{l_1}^2 {l_2}^2}\ \frac{{\hat\phi_p}^*}{{l'_1}^2 {l'_2}^2}\frac{{\hat\phi}_i}{{a_1}^2 {a_2}^2}\ \frac{{\hat\phi_p}^*}{{a'_1}^2 {a'_2}^2}\
\nonumber\\
&\times&T_2(l_2, a_2\to q_2,q'_2)\;T_2^*(l'_2, a'_2\to  q_2,q'_2)\;
T_1(l_1, a_1\to q_1,q'_1)\; T_1^*(l'_1,a'_1\to q_1,q'_1)\nonumber\\
&\times&\delta(L-l_1-l_2-F_3)\;\delta(L'-l'_1-l'_2-F_3)\nonumber\\&\times&\delta(N-a_1-a_2-F_1)\;\delta(N'-a'_1-a'_2-F_1)\;\nonumber\\
&\times&\delta(l_1+a_1-Q_1)\:\delta(l'_1+a'_1-Q_1)\;\delta(l_2+a_2-Q_2)\;\delta(l'_2+a'_2-Q_2)\nonumber\\
& \times&\prod_{i,j}d(\Omega_i/8) \;d^4a_i d^4a'_i d^4l_i d^4l'_i d^4 F_j\delta ({F_j}^2-{M_j}^2) \;d^4 Q_id{M_j}^2
\end{eqnarray}

Here and later on, we use the following notation: $\phi_i$ is the effective  vertex for one-parton emission, $\hat\phi_i$ the effective vertex for two-parton emission, $i=p$ when the nucleon is the projectile proton, $i=1,2$ labels the bound nucleons of the target nucleons. $T_i$ is the T-matrix for the parton scattering (the index $i$ labels the corresponding bound nucleon). The final
 state momenta are $q_i,\,q'_i$, while $Q_i=q_i+q'_i$ is the overall four-momentum of the final state $i$. The final directions are embodied in the angle
 $\Omega_i$, the factor $1/8$ relates the invariant relative phase space with the solid angle.  We use the fact that the momentum variables have large and small components, so the $plus$ components of $L,l_i,F_3$ are large and the corresponding $minus$ variables are small. On the contrary the $minus$ components of $N_i,a_i,F_i$ are large and the $plus$ components are small. The four-momenta of the produced particles can have both $plus$ and $minus$ large components. More explicitly: large means $\propto\sqrt s$, small means $\propto 1/\sqrt s$, the transverse variables are constant with respect to centre of mass energy $s$ and the general attitude, as in\cite{Calucci:2010wg}, is to integrate over the small components. \par
It is useful to introduce now the fractional $plus$ or $minus$ momenta in the following way:
 $$x_1={l_1}_+/L_+,\;x_2={l_2}_+/L_+,\;z_1={a_1}_-/N_-,\;z_2={a_2}_-/N_-$$
  $L$ is the four momentum of the free proton, $N$ the four momentum of the other nucleon, $D$  the four momentum of the Deuteron, $T$  the four momentum of the Tritium.
 The T-matrix amplitudes are related to the partonic cross section by:

 \begin{equation}
|T(l+a\to qq')|^2=(8\pi)^2(l_+a_-)d\hat\sigma_Q/d\Omega=
 (8\pi)^2xz (L_+N_-)d\hat\sigma_Q/d\Omega
 \end{equation}

Hadronization is not included.
The cross section is obtained from the discontinuity of the forward amplitude $(L=L',\;N=N')$, removing the overall four momentum conservation and dividing by the incoming flux $2s$.

\par
 If one of the colliding nucleons is bound, then we define the fractional momentum of the nucleon as $ Z=2N_-/D_-$, in the case of the Deuteron, and
$ Z=3N_-/T_-$, in the case of Tritium. It is useful to introduce also the variables $\bar x_i$, defined as $\bar x_i=2a_{i-}/D_-$, for the Deuteron, and $\bar x_i=3a_{i-}/T_-$, for the Tritium. The fractional momenta of partons
  with respect to the parent nucleons are thus
  $z_1=\bar x_1/(2- Z)$, $z_2=\bar x_2/ Z$, for the Deuteron and analogously for the  Tritium. In the expression of the flux $N_-$ is substituted by $D_-$ and $T_-$ respectively.

\begin{figure}[h]
\centering
\includegraphics[width=160mm]{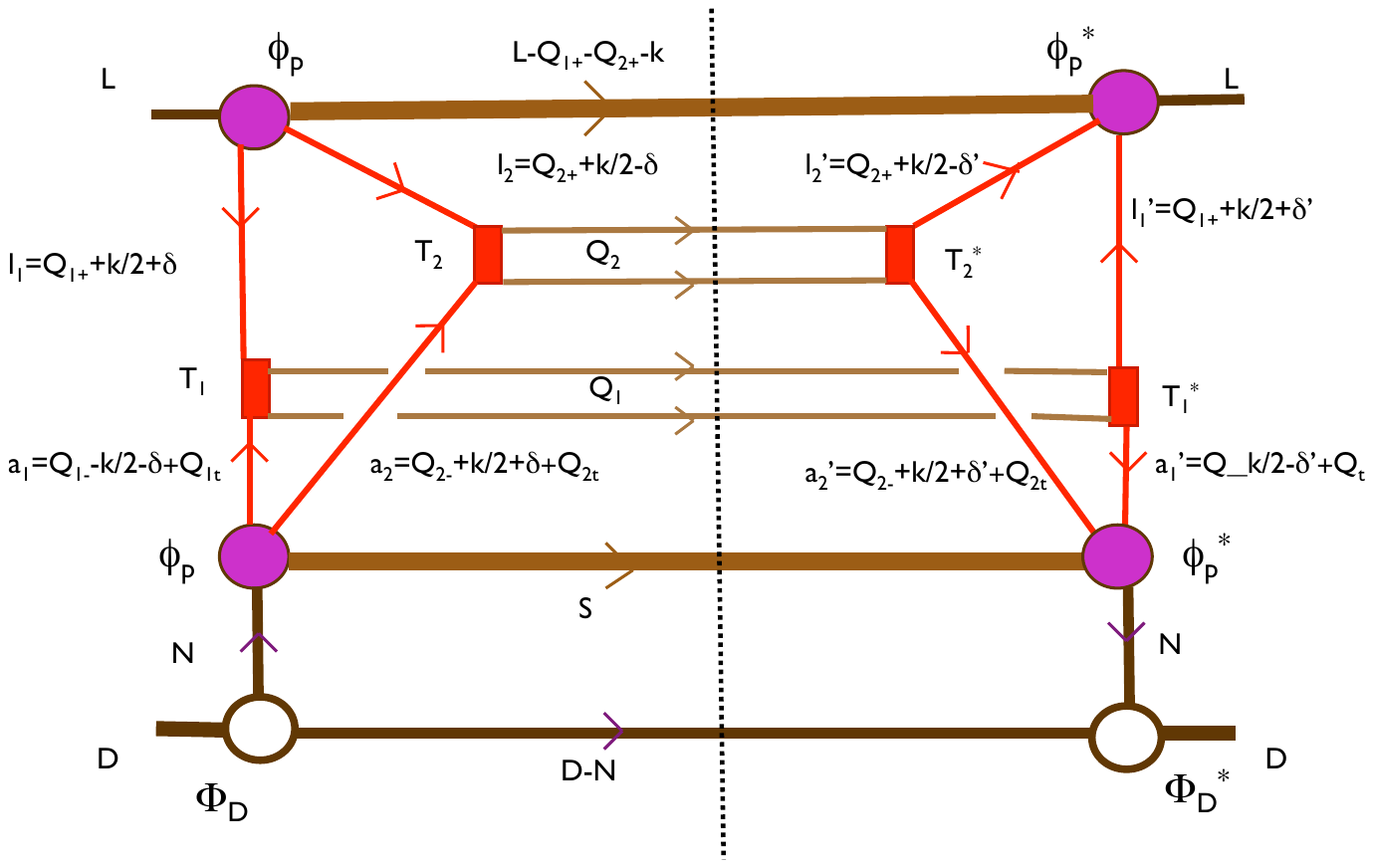}
\caption {Double parton scattering in $p\ D$ interactions. Only a single target nucleons interact with large momentun exchange} \label{10}
\end{figure}

                                                                   \par
If both interacting nucleons are free, one obviously has $L^2=m^2,\;N_1^2=m^2$.
In the case of a Deuteron with a spectator nucleon, the relevant discontinuity is

\begin{eqnarray}
 {\rm Disc} \;{\cal A}_{(2;0)} &&=\int dNdN'\;{\rm Disc}\;{\cal F}_2(L,L',D,D')
\delta((D-N)-(D'-N'))\cr
  \times\frac{\Phi_D(N)}{[N^2-m^2]}\frac{{\Phi_D}^*(N')}{[N'^2-m^2]}
 && \delta((D-N)^2-m^2)/(2\pi)^3
 \end{eqnarray}

\noindent
where the Deuteron effective vertex $\Phi_D$ is introduced and the condition $N_1^2=m^2$ is substituted by $D^2=D'^2=M^2_D$.\par
In the case of a $^3H$ or $^3He$ target with two spectator nucleons, the corresponding discontinuity is

\begin{eqnarray}
 {\rm Disc}\;{\cal B}_{(2,0,0)}&&=\int dNdN'dN_3{\rm Disc}\;{\cal F}_2(L,L',D,D')\delta(T-N-T'+N')\cr
 \times\frac{\Phi_T(N,N_3)}{[N^2-m^2]}\frac{{\Phi_T}^*(N',N_3)}
 {[N'^2-m^2]}&&\delta (N_3^2-m^2)\delta ((T-N-N_3)^2-m^2)/(2\pi)^6
\end{eqnarray}

\noindent
where the Tritium (or $^3He$) effective vertex $\Phi_{T}$ is introduced and the mass-shell condition is: $ T^2=M^2_T$.

 \par Following\cite{Calucci:2010wg}, we proceed by defining the amplitude for finding one or two partons in the projectile when the remnant of the parent nucleon has mass $M$. The integrated variable is $\lambda_-=\frac{1}{2}(l_1-l_2)_-$ and $M_{\bot}$ is the transverse mass.

\begin{eqnarray}
 \psi_{1,M}&=&\frac {\phi_p}{l^2}=
 \frac {\phi_p}{x[m^2-M_{\bot}^2/(1-x)]-l_{\bot}^2}\cr
 \psi_{2,M}&=&\frac{1}{\sqrt 2}\int\frac{\hat\phi_p}{l_1^2 l_2^2}\frac{d\lambda_-}{2\pi i}  =\frac {1}{\sqrt{2} L_-}\frac{\hat\phi}{{l_1}_{\bot}^2 x_2+{l_2}_{\bot}^2 x_1-
 x_1,x_2[m^2-M_{\bot}^2/(1-x_1-x_2)]}\;.
  \end{eqnarray}

The one-parton and two-parton amplitudes in the bound nucleon
are defined in the same way. The only difference is that in the case of the bound nucleon one needs to replace $m^2$ with $m^2+N_{\bot}^2$. The covariant amplitude for finding a nucleon in the Deuteron is defined in an analogous way:

\begin{eqnarray}
\frac{1}{\sqrt 2} \int\frac {\Phi_D}{[(D-N)^2-m^2]\cdot [N^2-m^2]}\frac{d N_+}{2\pi i}
 &=&\frac{1}{\sqrt 2}\frac {1}{N_-}\frac{\Phi_D}{[(D-N)^2-m^2]}\Big|_{N^2=m^2}
 =\frac{\Psi_D(N_-)}{N_-}\cr
=\frac{1}{\sqrt 2}\frac {1}{(D-N)_-}\frac{\Phi_D}{[N^2-m^2]}\Big|_{(D-N)^2=m^2}
 &=&\frac{\Psi_D((D-N)_-)}{(D-N)_-}
 \end{eqnarray}

\noindent
with the definition $\Psi_D(N_-)/N_-=\Psi_D((D-N)_-)/(D-N)_-$.
 We have also

 $$\frac{\Psi_D(N_-)}{N_-}=\frac{\Phi}{\sqrt
 2}\frac{1}{D_-[M_D^2Z_1Z_2/4-m_{\bot}^2]}\qquad Z_1+Z_2=2\quad
m_{\bot}^2=m^2+N^2_{\bot} $$

\noindent
As a function of $Z_1$, the amplitude $\Psi_D(N_-)/N_-$ has a maximum for $Z_1=1$, which obviously implies also $Z_2=1$.

\par
  Finally we define the covariant amplitude for finding two nucleons in the Tritium

\begin{eqnarray}
\frac{1}{2(2\pi i)^2}& &\int\frac {\Phi_T}{[N_1^2-m^2]\cdot [N_2^2-m^2]\cdot [N_3^3-m^2]}
\prod_j d N_{j+}\delta(T_+ -\sum_j N_{j+})\cr
&=&\frac{9}{2T_-^2}\frac {\Phi_T}{M_T^2Z_1Z_2Z_3/3-m_{\bot,1}^2 Z_2Z_3-m_{\bot,2}^2Z_3Z_1-m_{\bot,3}^2Z_1Z_2}
\end{eqnarray}

The expression is evidently symmetrical in $(1,2,3)$ and can thus be indentifyed with
$\Psi_T({N_1}_-,{N_2}_-)/({N_1}_-{N_2}_-)$, or with $\Psi_T({N_2}_-,{N_3}_-)/({N_2}_-{N_3}_-)$, or with $\Psi_T({N_3}_-,{N_1}_-)/({N_3}_-{N_1}_-)$. The corresponding Tritium amplitudes have also a maximum in $Z_i$, where $\sum_i Z_i=3$, close although not strictly equal to one. A more detailed inspection shows that the lack of strict equality is due to the possible difference in the three transverse momenta. By defining $\gamma_i=3m_{i\bot}^2/M_T^2-1/3$, we find that, at the first order in $\gamma_i$, the Tritium amplitude is maximized for $Z_i=1-\sum_j\gamma_j/6+\gamma_i/2$. 

\par
We proceed now with the integration on the transverse variables, in the frame where the external transverse momenta $L_{\bot}$, and $D_{\bot}$, are equal to zero. We take the two-dimensional Fourier transforms ( $b_i$ is conjugated to $a_i$, $B_j$ to $N_j$, $\beta_i$ to $l_i$, all the variables are two-dimensional vectors).

  \begin{eqnarray}
  \psi_1&=&(2\pi)^{-1}\int\tilde\psi_1\exp [il b]db\quad
 \psi_2=(2\pi)^{-2}\int\tilde\psi_2 (b_1,b_2)\exp [il_1 b_1+il_2b_2]db_1db_2\cr
 \Psi&=&(2\pi)^{-1}\int\tilde\Psi (B)\exp [iN B] dB
 \end{eqnarray}

\noindent
and analogously for the complex conjugated functions, with the variables
  $b'_1,b'_2,B'$.\par
  The integration over the transverse-momentum variables gives the diagonal
  property $b_1=b'_1$ and so on. Moreover one obtains the geometrical
  relation: $b_1-b_2=\beta_1-\beta_2$.

  \par
 The one-body and two-body parton densities are defined by the following integrals on the invariant mass of the residual hadron fragments:
\begin{eqnarray}
\Gamma (z;b)&=&\frac{1}{2(2\pi)^3}\int |\tilde\psi_M(z;b)|^2\frac{z}{1-z}dM^2\cr
\Gamma (x_1,x_2;b_1,b_2)&=&\frac{1}{2(2\pi)^6}\int|\tilde\psi_M(x_1,x_2;b_1,b_2)|^2\frac{x_1,x_2}{1-x_1-x_2}L_+^2dM^2
 \end{eqnarray}

The residual dependence on $q_{1\bot}$, $ q'_{1\bot}$ is transformed into an angular dependence on $\Omega_1,\;\Omega_2$. \par
 As an effect of the nucleon motion,  $|\tilde\Psi_D( z;b)|^2$ is coupled to the interactions by the integration on the fractional momentum $Z$, while the integration on the transverse variable $B$ is decoupled from the other transverse variables. So the cross section is readily expressed as\par

\begin{eqnarray}
&\sigma^{pD}_{2,1}&=\frac{2}{(2\pi)^3}\int\Gamma(x_1,x_2;b_1,b_2)\frac{d\sigma(x_1x_1')}{d\Omega_1}
\frac{d\sigma(x_2x_2')}{d\Omega_2}\Gamma(x_1'/ Z,x_2'/ Z;\beta_1,\beta_2)\cr
&&\qquad |\tilde\Psi_D(2-Z;B)|^2/(2-Z)dB\,db_1\, db_2\,d\beta_1\,d\beta_2\,\delta(b_1-b_2-\beta_1+\beta_2)\cr
&&\qquad dx_1dx_2dx'_1dx'_2\,d Z\,d\Omega_1\,d\Omega_2
\end{eqnarray}

Note that the main dependence on $Z_i$, for $Z_i$ close to one,  is due to the nuclear wave function $\Psi_D$, while the error, due to the approximation $Z=1$ in the parton distribution $\Gamma$, is less important. The expression above can thus be slightly transformed as follows.

From the properties of $\Psi$ we have:

 $$|\Psi_D(2-Z)|^2/(2-Z)=[1+(1-Z)]|\Psi_D(Z)|^2/Z^2$$
 
\noindent
 the second addendum is odd for the substitution $Z\to (2-Z)$ so that the integration in $Z$ which runs from 0 to 2 gives zero and the cross section is more conveniently expressed by:

\begin{eqnarray}
&\sigma^{pD}_{2,1}&=\frac{2}{(2\pi)^3}\int\Gamma(x_1,x_2;b_1,b_2)\frac{d\sigma(x_1x_1')}{d\Omega_1}
\frac{d\sigma(x_2x_2')}{d\Omega_2}\Gamma(x_1'/ Z,x_2'/ Z;\beta_1,\beta_2)\cr
&&\qquad |\tilde\Psi_D(Z;B)|^2/Z^2dB\,db_1\, db_2\,d\beta_1\,d\beta_2\,\delta(b_1-b_2-\beta_1+\beta_2)\cr
&&\qquad dx_1dx_2dx'_1dx'_2\,d Z\,d\Omega_1\,d\Omega_2
\end{eqnarray}

\subsection {Two bound nucleons interact with large momentum transfer}

In collisions of protons with $D$ or $^3H/^3He$, the presence of the nuclear wave function induces the presence of two kinds of contributions: "diagonal terms" in direct correspondence with the processes taking place when the nucleons are free and a number of nondiagonal or "interference" terms, which are due to the presence of the nuclear wave function$^1$.\footnotetext[1]{The discussion in\cite{Calucci:2010wg} is limited to the cases where the contribution of the interference term is not relevant.}

\begin{figure}[h]
\centering
\includegraphics[width=160mm]{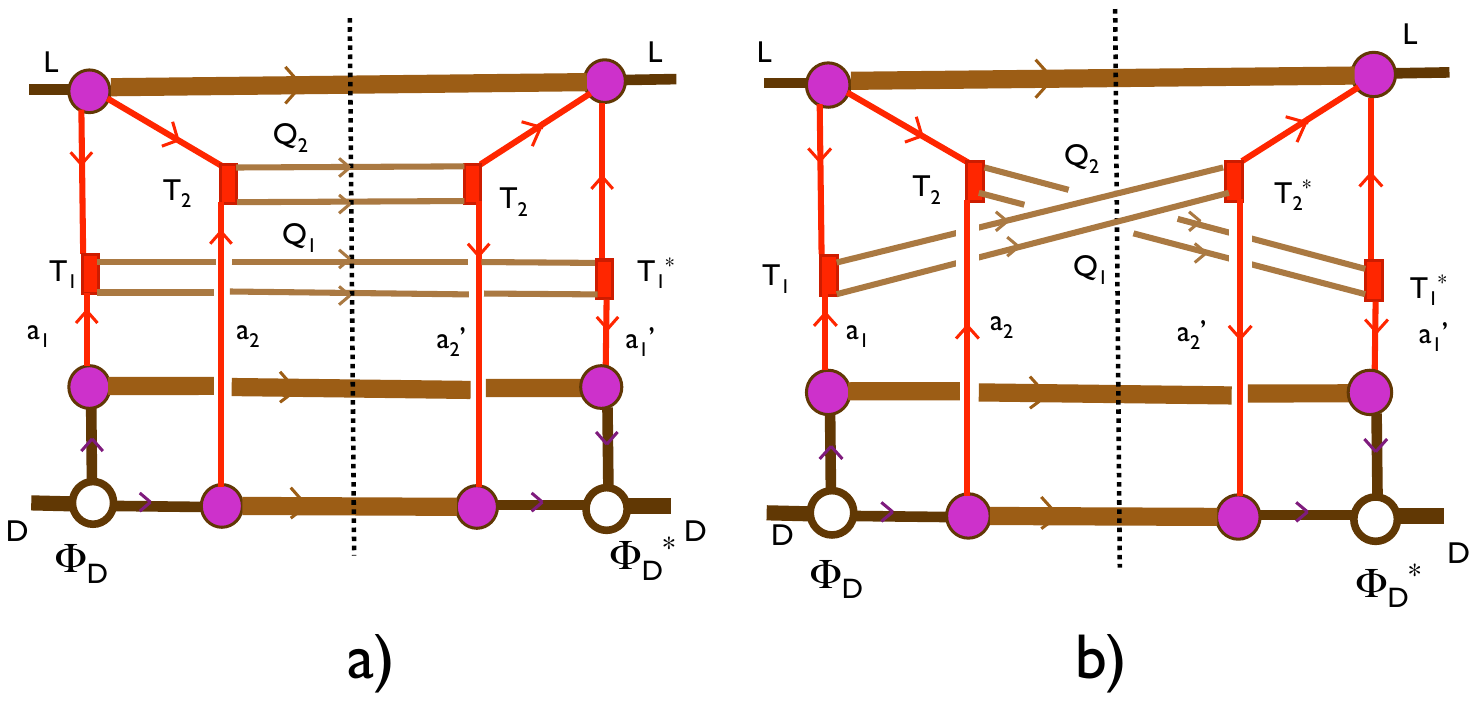}
\caption {Double parton scattering in $p\ D$ interactions. Both target nucleons interact with large momentun exchange} \label{11}
\end{figure}

 \par
  The simplest case, where diagonal and nondiagonal terms appear, is the double scattering on a Deuteron affecting both bound nucleons. In this case the "diagonal" discontinuity has the already given form:
\begin{eqnarray}
{\rm Disc}{\cal A}_d^{(2)}&=&\frac{1}{(2\pi)^{21}}\int\frac{{\hat\phi}_p}{{l_1}^2 {l_2}^2}\ \frac{{\hat\phi_p}^*}{{l'_1}^2 {l'_2}^2}\ \frac{\phi_p}{a_1^2}\;
\frac{\phi_p^*}{{a'_1}^2}\;\frac{\phi_n}{{a_2}^2}\;\frac{\phi_n^*}{{a'_2}^2}\nonumber\\
&\times&T_2(l_2, a_2\to q_2,q'_2)\;T_2^*(l'_2, a'_2\to q_2,q'_2)\;
T_1(l_1, a_1\to q_1,q'_1)\; T_1^*(l'_1,a'_1\to q_1,q'_1)\nonumber\\
&\times&\frac{\Phi_D(D;N)}{[(D-N)^2-m^2][N^2-m^2]}\frac{\Phi_D^*(D;N')}{[(D-N')^2-m^2][N'^2-m^2]}\nonumber\\
&\times&\delta(L-l_1-l_2-F_3)\;\delta(L-l'_1-l'_2-F_3)\nonumber\\&\times&\delta(N-a_2-F_2)\;\delta(N'-a'_2-F_2)\;\delta(D-N-a_1-F_1)\;\delta(D-N'-a'_1-F_1)\nonumber\\&\times&\delta(l_1+a_1-Q_1)\:\delta(l'_1+a'_1-Q_1)\;\delta(l_2+a_2-Q_2)\;\delta(l'_2+a'_2-Q_2)\nonumber\\
& \times&\prod_{i,j}d(\Omega_i/8) \;d^4a_i d^4a'_i d^4l_i d^4l'_i d^4 F_j\delta ({F_j}^2-{M_j}^2) \;d^4N d^4N' d^4 Q_id{M_j}^2
\end{eqnarray}

\noindent
whereas the interference term has the form:

\begin{eqnarray}
{\rm Disc}{\cal A}_i^{(2)}&=&\frac{1}{(2\pi)^{21}}\int\frac{{\hat\phi}_p}{{l_1}^2 {l_2}^2}\ \frac{{\hat\phi_p}^*}{{l'_1}^2 {l'_2}^2}\ \frac{\phi_p}{a_1^2}\;
\frac{\phi_p^*}{{a'_1}^2}\;\frac{\phi_n}{{a_2}^2}\;\frac{\phi_n^*}{{a'_2}^2}\nonumber\\
&\times&T_2(l_2, a_2\to q_2,q'_2)\;T_1^*(l'_1, a'_1\to q_2,q'_2)\;
T_1(l_1, a_1\to q_1,q'_1)\; T_2^*(l'_2,a'_2\to q_2,q'_2)\nonumber\\
&\times&\frac{\Phi_D(D;N)}{[(D-N)^2-m^2][N^2-m^2]}\frac{\Phi_D^*(D;N')}{[(D-N')^2-m^2][N'^2-m^2]}\nonumber\\
&\times&\delta(L-l_1-l_2-F_3)\;\delta(L-l'_1-l'_2-F_3)\nonumber\\&\times&\delta(N-a_2-F_2)\;\delta(N'-a'_1-F_2)\;\delta(D-N-a_1-F_1)\;\delta(D-N'-a'_2-F_1)\nonumber\\&\times&\delta(l_1+a_1-Q_1)\:\delta(l'_1+a'_1-Q_1)\;\delta(l_2+a_2-Q_2)\;\delta(l'_2+a'_2-Q_2)\nonumber\\
& \times&\prod_{i,j}d(\Omega_i/8) \;d^4a_i d^4a'_i d^4l_i d^4l'_i d^4 F_j\delta ({F_j}^2-{M_j}^2) \;d^4N d^4N' d^4 Q_id{M_j}^2
\end{eqnarray}

The diagonal term was already elaborated in\cite{Calucci:2010wg}, so we are interested in the differences
between the two cases. In the diagonal case the conservation of the large components of momenta implies that they are equal on the two sides of the diagram $l_+=l'_+,\;a_-=a'_-,\;N_-=N'_-$; the transverse variables become diagonal through the Fourier transformation. In this way the whole expression of the cross section is expressible in terms of densities $i.e.$ square of the partonic wave function and square of the nuclear wave function.
The cross section is thus expressed again through the one-body and two-body partonic densities:
$\Gamma(z;b), \quad \Gamma(x_1,x_2;\beta_1,\beta_2)$, obtained from the effective vertices $\phi,\;\hat\phi$.
The nuclear density is simply given by $|\Psi(Z;B)|^2$.

\par
 In the interference case the conservation of the large components of momenta still implies the equality on the two sides of the diagram $l_+=l'_+,\;a_-=a'_-$ but for the nuclear variables $(N-a_{2})_-=(N'-a_{1})_-$, moreover the transverse variable $b_{\bot}$, conjugated to $a_{\bot}$, does not become diagonal through the Fourier transformation.
 The interference term cannot be expressed only in terms of partonic densities, we need to introduce a more complicated expression:

  \begin{eqnarray}
 && W_1(Z,Z';\bar x_1,\bar x_2;b_1,b_2,B)=\frac {1}{4(2\pi)^6} \int dM_1^2dM_2^2
 \frac {\bar x_1\bar x_2}{(Z-\bar x_1)(2-Z'-\bar x_2)}        \cr
  && \times \psi_{M_1}(\bar x_1/Z;b_1) \psi^*_{M_2}(\bar x_1/(2-Z');b_1-B)\psi_{M_2}(\bar x_2/(2-Z);b_2) \psi^*_{M_1}(\bar x_2/Z';b_2+B)
 \label{eq:W1}
   \end{eqnarray}

  The previous relation for the nuclear variables can be written also $Z-Z'=\bar x_2-\bar x_1$. It is possible to factor the expression $W$ symmetrically into two parts
$W_1=H(Z,Z';\bar x_1,\bar x_2;b_1,b_2+B)\times H(2-Z,2-Z';\bar x_2,\bar x_1;b_2,b_1-B)$ with
  $$ H(Z,Z';\bar x_1,\bar x_2;b_1,b_2)=\frac {1}{2(2\pi)^3} \int dM_1^2
  \frac {\sqrt{\bar x_1\bar x_2}}{\sqrt{(Z-\bar x_1)(2-Z'-\bar x_2)}}
  \psi_{M_1}(\bar x_1/Z;b_1)\psi^*_{M_1}(\bar x_2/Z';b_2)\;. $$
  The part of the cross section coming from the direct term is (as given in\cite{Calucci:2010wg}) by

\begin{eqnarray}
\sigma^{pD}_{2,2}\big|_d&=&\frac{1}{(2\pi)^3}\int\Gamma(x_1,x_2;\beta_1,\beta_2)
\frac{d\hat\sigma(x_1,\bar x_1)}{d\Omega_1}
\frac{d\hat\sigma(x_2,\bar x_2)}{d\Omega_2}
\Gamma(\bar x_1/ Z;b_1)\Gamma(\bar x_2/(2- Z);b_2)\cr
&\times &|\tilde\Psi_D( Z;B)|^2 dB\,db_1\, db_2\,d\beta_1\,d\beta_2\delta(B-b_1 +b_2-\beta_1+\beta_2)\cr
&\times &Z^{-2}\,dx_1dx_2d\bar x_1d\bar x_2\,d Z\,d\Omega_1\,d\Omega_2
\end{eqnarray}

In fig.3 we show the corresponding configuration in transverse space.

\begin{figure}[h]
\centering
\includegraphics[width=110mm]{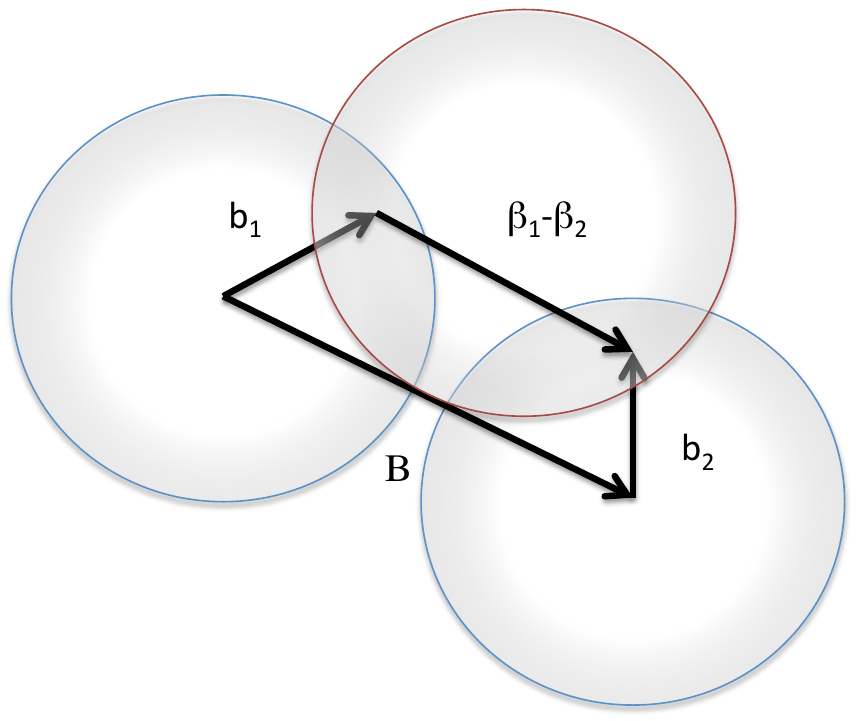}
\caption {$p\ D$ interactions with two target nucleons involved. Configurations in transverse space of the diagonal term} \label{11D}
\end{figure}

   The part of the cross section coming from the interference term is

\begin{eqnarray}
\sigma^{pD}_{2,2}\big|_i&=&\frac{1}{(2\pi)^3}\int\Gamma(x_1,x_2;\beta_1,\beta_2)\frac{d\hat\sigma(x_1,\bar x_1)}{d\Omega_1}
\frac{d\hat\sigma(x_2,\bar x_2)}{d\Omega_2}W_1(Z,Z';\bar x_1,\bar x_2;b_1,b_2,B)\cr
& \times&\tilde\Psi_D( Z;B) \tilde\Psi_D^*( Z';b)\big[ZZ'\big]^{-1}
\delta(B-b_1 +b_2-\beta_1+\beta_2)\delta(Z-Z'-\bar x_1+\bar x_2)\cr
&\times&d\Omega_1\,d\Omega_2\,dB\,\,db_1\, db_2\,d\beta_1\, d\beta_2\,dx_1dx_2d\bar x_1d\bar x_2\,dZdZ'
\end{eqnarray}

In fig.4 we show the configurations in transverse space of the two interfering amplitudes.

\begin{figure}[h]
\centering
\includegraphics[width=120mm]{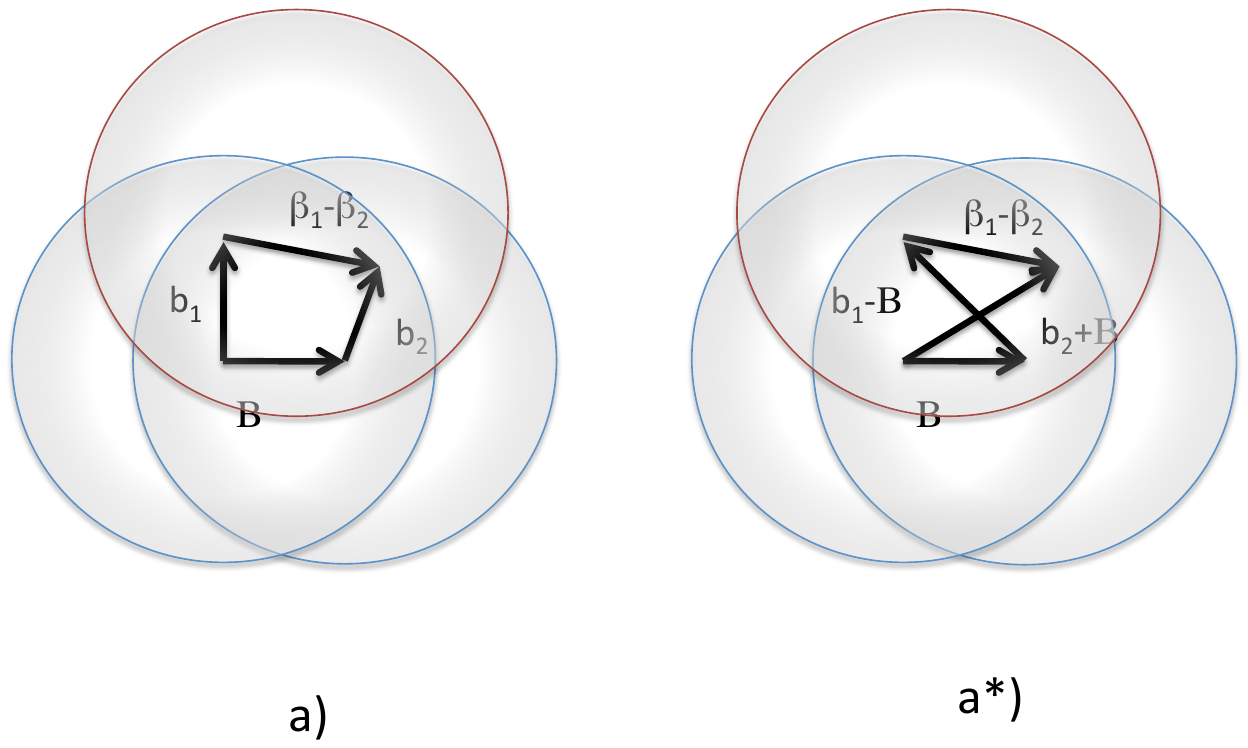}
\caption {$p\ D$ interactions with two target nucleons involved. Configurations in transverse space of the two interfering amplitudes. Both configurations generate the same partonic initial state.} \label{11O}
\end{figure}

A comparison between the two expressions: As far as the longitudinal variables are concerned the interference term requires the nuclear wave function to be taken at different values of $Z$, so it is depressed with respect to the diagional term. It must be taken into account that the width of the nuclear wave function is determined by the binding energy while the mismatch of the $Z$ terms depend on the difference in fractional momentum of the partons. At large total energies (beyond the TeV) the values of $\bar x$ can be small, still mantaining the process within the limits of perturbative  dynamics, so the depression is not necessarily very strong. For what concerns the transverse variables the possible depression is given by the ratio between the hadronic size (as seen in the hard QCD processes) and the nuclear size. The first scale sets the size of the variables $b_i, \beta_i$ the second characterizes the size of $B$. After integrating over $B$ with the $\delta-$function, one has $B=(b_1-b_2+\beta_1-\beta_2)$. A simplified form for the interference term is thus obtained when neglecting the hadronic size as compared with the nuclear size by setting $B=0$ in nuclear wave function $\Psi$.
Note that the effect depends on the transverse degrees of freedom, while the dependence on the total energy is weak, once the region of perturbative dynamics is reached. A quantitative, but model dependent discussion of this feature can be found in Appendix B.

\par Conversely the denominators $Z, Z'$ could be set equal to $1$ in the factors $\Gamma$ since the range of variation of $\bar x_i$ is large, as compared with the variation of $Z$, allowed by the nuclear function $\Psi$.
From the previous treatment we learned that the presence of a spectator has a modest influence on the process, so the double scattering on a Tritium has only minor differences in comparison with the scattering on a Deuteron. The property holds also in the next cases to be considered, with the exception of those processes, which are possible in presence on Tritium but not on a Deuteron.

   \section {Triple scattering on Deuteron or Tritium}
  \subsection {Only one bound nucleon interacts with large momentum transfer}

The analytical expression for the hard scattering, where one of the component nucleons interacts three times and there are one (Deuteron) or two (Tritium) spectators, is again related to the amplitude ${\cal F}_3$ for the triple hard scattering between two free nucleons:

 \begin{eqnarray}
 Disc \;{\cal F}_3&=&
{1\over {(2\pi)^{32}}}\int {{\check\phi_p}\over{l_1^2 l_2^2l_3^2}}
{{{\check\phi_p}^*}\over{{l'}_1^2 {l'}_2^2{l'}_3^2}}
{{\check\phi_1}\over{a_1^2 a_2^2 a_3^2}}
{{{\check\phi_1}^*}\over{{a'}_1^2 {a'}_2^2 {a'}_3^2}} T_1(l_1a_1\to q_1 q'_1) {T_1}^*({l'}_1{a'}_1\to q_1q_2)\cr
&\times& T_2(l_2a_2\to q_2q'_2) {T_2}^*({l'}_2{a'}_2\to q_2q'_2)T_3(l_3a_3\to q_3q'_3) {T_1}^*({l'}_3{a'}_3\to q_3q'_3)\cr
&\times& \delta(L-\sum l-F_4)  \delta(L-\sum {l'}-F_4)\delta(N_1-\sum a-F_1)\delta
({N'}_1-\sum {a'}-F_1)\cr
&\times&\delta(l_1+a_1-Q_1)\delta({l'}_1+{a'}_1-Q_1)\delta(l_2+a_2-Q_2)\delta({l'}_2+{a'}_2-Q_2)
\cr
&\times&\delta(l_3+a_3-Q_3)\delta({l'}_3+{a'}_3-Q_3)\delta(N_1-{N'}_1)\cr
&\times& \prod_{i,j}d(\Omega_i/8) \;d^4a_i d^4a'_i d^4l_i d^4l'_i d^4 F_j\delta ({F_j}^2-{M_j}^2) \;d^4N_1 d^4 Q_id{M_j}^2
\end{eqnarray}

\begin{figure}[h]
\centering
\includegraphics[width=160mm]{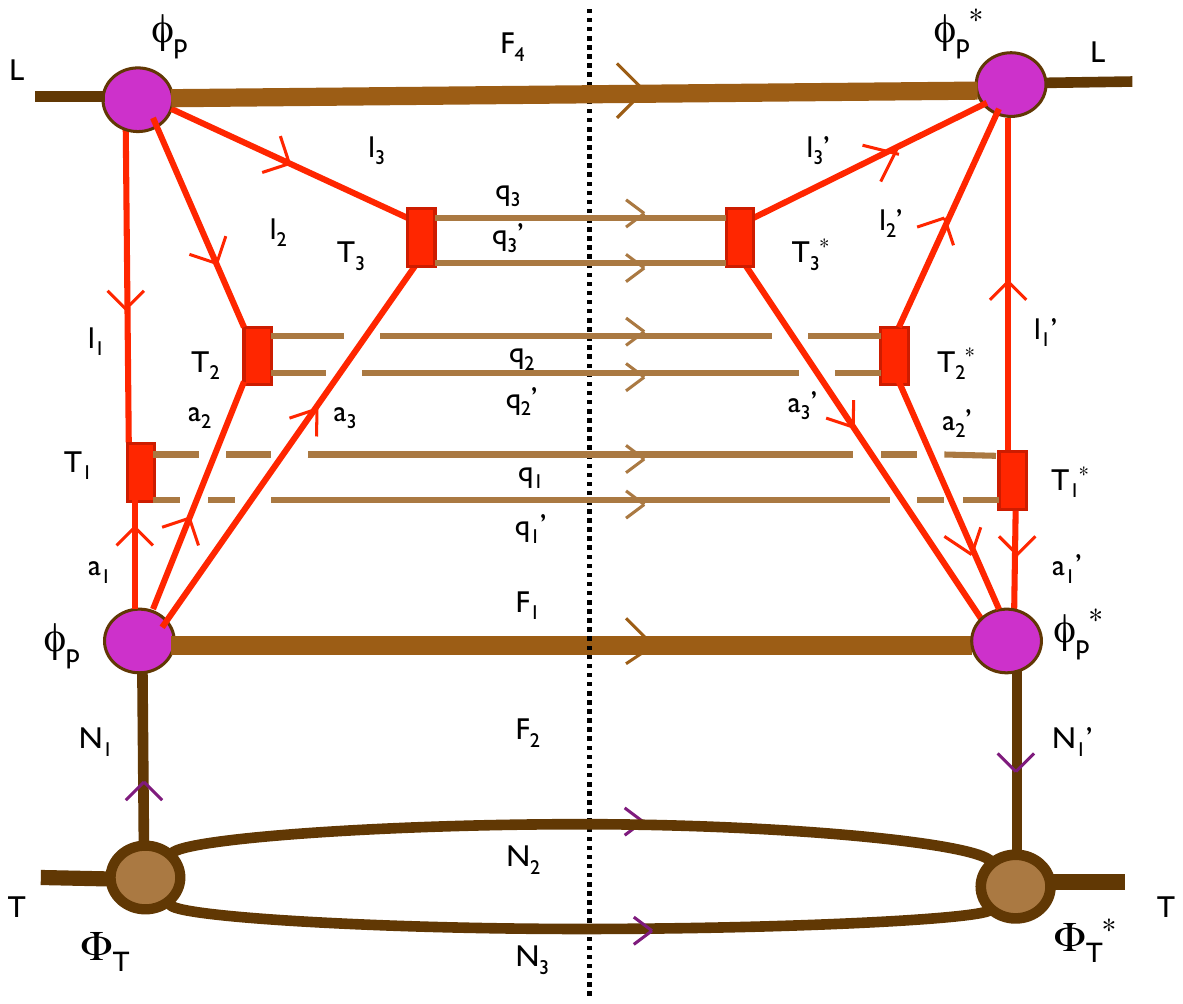}
\caption {Triple parton scattering in $p\ T$ or $p\ ^3He$ interactions. Only a single target nucleons interacts with large momentun exchange} \label{300}
\end{figure}

\noindent
Analogously to the case of double scattering,
also for the calculations of ${\cal F}_3$ we use the property that, in the regime of interest, the momentum variables have large and small components; the term $\check\phi$ represents the vertex for emission of three partons from a nucleon.\par
The factor:
 $1/(l_1^2\,l_2^2\,l_3^2)$ is thus integrated in ${l_1}_-$ and ${l_2}_-$ independently from the rest of the diagram, in fact in all other conservation relations, the small terms ${l_1}_-$ and ${l_2}_-$ enter together with large components $e.g.$ of $a_i$ and can thus be neglected. Implementing the conservation $(L-F_4)=l_1+l_2+l_3$, the integration gives
$$\psi_{3,M}=\frac{1}{(2\pi i)^2 }\int\frac{1}{2} d{l_1}_-d{l_2}_-\frac{\check\phi}{l_1^2\,l_2^2\,l_3^2}=-\frac{1}{2}
\frac{\check\phi}{{l_1}_+{l_2}_+{l_3}_+(L-F_4)_- -{l_1}_+{l_2}_+{l_3}_{\bot}^2-{l_2}_+{l_3}_+{l_1}_{\bot}^2-{l_3}_+{l_1}_+{l_2}_{\bot}^2}$$
We use here also the fractional momenta: $x_i={l_i}_+/L_+\simeq {Q_i}_+/L_+$ and, through the conservation for the $plus$ components, we obtain:

\begin {equation}
\psi_{3,M}(l)={{1}\over{2{L_+}^2}}
{\check\phi\over{{x_1}{x_2}{l_3}_{\bot}^2+{x_2}{x_3}{l_1}_{\bot}^2+{x_3}{x_1}{l_2}_{\bot}^2+[M^2_{\bot}/(1-\sum x_i)-m^2]x_1,x_2,x_3}}\;.
\end{equation}

\noindent
$M_{\bot}$ is the transverse mass of the remnants:  ${M_{\bot}}^2={M_F}^2+{F_{\bot}}^2,\; F_{\bot}=-\sum {l_i}_{\bot} $\par
The factor $\check\phi/(a_1^2\,a_2^2\,a_3^2)$ can be integrated in the same way. One needs only to keep into account that we are interested in a situation where $N_1$ is a generic time-like four vector, with positive energy, no longer subjected to the condition $N_1^2=m^2$. Even when $N_i$ enters in a wave function, it is still almost on shell, as we are dealing with a weakly bound systems (it has nevertheless a transverse momentum $N_{i\bot}$).
 Let $F$ be the four-momentum of the remnants; then in strict analogy with the previous result we get:

 \begin {equation}
\psi_{3,M}(a)={{1}\over{2{N_-}^2}}
{{\check\phi}\over{{z_1}{z_2}{a_3}_{\bot}^2+{z_2}{z_3}{a_1}_{\bot}^2+{z_3}{z_1}{a_2}_{\bot}^2+
[M^2_{\bot}/(1-\sum z_i)-m_N^2]z_1z_2z_3}}\;.
\end{equation}

\noindent
 $M_{\bot}$ is again the transverse mass of the remnants, here the conservation of the $minus$ component is used.
The integration on the remnants $F$ can be treated as in the double scattering case:
$\int d^4F\delta (F^2-M^2)dM^2=\int dF_{\pm}/F_{\pm}d^2 F_{\bot}$. The longitudinal integration is then performed by means of the $\delta$-functions with the results: $1/{F_1}_-=1/[N_-(1-\sum z)]\quad 1/{F_4}_+=1/[L_+(1-\sum x)].$\par
Looking at the cut graph in Fig.1, we see that the equality of $N_2$ and, in case, of $N_3$ forces also $N_1$ to be the same on the right and on the left hand side of the diagram. Concerning the produced pairs, a quick inspection to the kinematics shows that the component $Q_+$ comes from $l_+$ and the component $Q_-$ comes from $a_-$, which, neglecting terms of order $1/\sqrt s$, imply ${l_i}_+={{l'}_i}_+$ and ${a_i}_-={{a'}_i}_-$. In terms of the fractional momenta $x_i=x'_i\;z_i=z'_i$ and
according with the definitions 

$$\prod dl_+da_-= (L_+ {N}_-)^3\prod dx dz\;.$$

Using the already defined $\psi_3(l)\,\psi_3(a)$
and analogously for the factors depending on ${l'}_i,{a'}_i$, we perform a Fourier transform on the transverse momenta; $\beta$ is the conjugated of $l$ and $b$ is the conjugated of $a$.\par
The subsequent integrations involve the nuclear variables. The longitudinal variables $Z$ are common to both sides of the cut diagram, the transverse variables are different.
 In the case of the Deuteron one has to integrate over one spectator (which is on shell) so we have integrations in $d Z_i/ Z_i$ and in ${dN_i}_{\bot}$, note that ${N_1}_{\bot}=-{N_2}_{\bot}$ and  $ Z_1=2- Z_2$.
 In the Tritium ($^3He$) case we have two transverse variables and the longitudinal integration, which may be expressed as:
 $\int \delta( Z_1+ Z_2+ Z_3-3)d Z_1 d Z_2 / Z_2 d Z_3/ Z_3$.
 In conclusion in both cases only one nuclear variable survives, and we need to perform the Fourier transform of the transverse components: we call the corresponding coordinates $B, B'$. \par
 The integrations which take care of the conservation conditions give two kinds of results:
the diagonalization in the impact parameters given as: $B=B',\;\beta_i=\beta'_i,\;b_i=b'_i$ and the geometrical conditions : $\beta_1-\beta_2=b_1-b_2,\;\beta_2-\beta_3=b_2-b_3$, which imply also $\beta_3-\beta_1=b_3-b_1$.

\par
 More in detail, the derivation of the condition on the transverse variables is obtained as follows (to simplify the notation, the transverse index, like $l_{\bot}$, is everywhere understood):\par
 The Fourier transform of the wave function is $\psi_3(l_1,l_2,l_3)=\int \tilde\psi_3 (\beta_i)\exp [i\sum l_i\beta_i]\prod d\beta$ and analogously for the functions in $a$ and for the conjugated.
 The conservation relations involving the produced pairs, integrated over
 the final states, give the two-dimensional constraints
 $$\delta(l_i+a_i-{l'}_i-{a'}_i)={1\over{(2\pi)^2}}\int d\theta_i \exp[i\theta_i(l_i+a_i-{l'}_i-{a'}_i)]$$
  The conservation between the incoming momenta and the momenta of the remnants gives a factor $\delta (\sum l-\sum {l'})$ and similar for the $a_i$, the sum is redundant, it is already contained in the previous relations, the difference gives a new condition $\delta (\sum l-\sum {l'}-\sum a+\sum {a'})$.
 In exponential form:
 $${1\over{(2\pi)^2}} \int d\zeta \exp [-i\zeta(\sum l-\sum {l'}-\sum a+\sum {a'})]\;. $$
 The integrations over the internal variables $l,{l'},a,{a'}$ give;
 
  $$\delta(\beta_i+\theta_i-\zeta),\quad \delta(\beta'_i+\theta_i-\zeta), \quad \delta(b_i+\theta_i-\zeta), \quad \delta(b'_i+\theta_i-\zeta)$$
 
 \noindent
  which in turn implies: $ \beta_i=\beta'_i,\quad b_i=b'_i$. One thus obtains the diagonalization in the impact parameter. Moreover one is left with $\delta(\beta_i-b_i-2\zeta)$, which represents a geometrical constraint: the difference $\beta_i-b_i$
 is independent of the index $i$.

 \par
 In analogy with the previous definitions\cite{Paver:1982yp}\cite{Calucci:2010wg}, the three-body densities are defined reabsorbing partially the factors $L_+,\; N_-$

 \begin{eqnarray}
 \Gamma(x_1,x_2,x_3;\beta_1,\beta_2,\beta_3)&=&{1\over{2(2\pi)^9}}\int
 |\tilde\psi(x_1,x_2,x_3;\beta_1,\beta_2,\beta_3)|^2{{x_1,x_2,x_3}\over{1-x_1-x_2-x_3}}{L_+}^4dM_4^2 \cr\cr
 \Gamma(z_1,z_2,z_3;b_1,b_2,b_3)&=&{1\over{2\,(2\pi)^9}}\int
 |\tilde\psi(z_1,z_2,z_3;b_1,b_2,b_3)|^2{{z_1z_2z_3}\over{1-z_1-z_2-z_3}}{N_-}^4dM_1^2
 \end{eqnarray}

 In this way the densities $\Gamma$ are such that they are neither vanishing nor growing indefinitely when $L_+,N_-\to \infty.$
The last expression could be recast in terms of the external variables $\bar x_i$ as

\begin{eqnarray}
 \Gamma(\bar x_1/ Z,\bar x_2/ Z,\bar x_3/ Z;b_1,b_2,b_3)&&\\
 ={ Z^2}{1\over{2^5(2\pi)^9}}\int&&
|\tilde\psi(\bar x_1/ Z,\bar x_2/ Z,\bar x_3/ Z;b_1,b_2,b_3)|^2{{\bar x_1\bar x_2\bar x_3}\over{ Z-\bar x_1-\bar x_2-\bar x_3}}{D_-}^4dM_1^2\nonumber  \end{eqnarray}

The triple scattering cross section on a Deuteron, which describes one of the bound nucleons suffering three hard collisions, while the other is a spectator, it is hence given by:

\begin{eqnarray}
&&\sigma^{pD}_{3,1}=\frac{2}{(2\pi)^3}\int
\Gamma(x_1,x_2,x_3;\beta_1,\beta_2,\beta_3)\Gamma(x_1'/ Z,x_2'/ Z,x_3'/ Z;b_1,b_2,b_3) \frac{d\sigma}{d\Omega_1}
\frac{d\sigma}{d\Omega_2}\frac{d\sigma}{d\Omega_3}\cr
&&\qquad |\tilde\Psi_D( Z;B)|^2 Z^{-2}dB\,db_1\, db_2\,db_3\,d\beta_1\,d\beta_2\,d\beta_3\,
\delta(b_1-b_2-\beta_1+\beta_2)\delta(b_1-b_3-\beta_1+\beta_3)\cr
&&\qquad dx_1dx_2dx_3dx'_1dx'_2dx'_3
\,d Z\,d\Omega_1\,d\Omega_2\,d\Omega_3\ .
\end{eqnarray}

When only a single target nucleon interacts with large momentum exchange, nuclear dynamics thus takes completely care of the difference between Deuteron and Tritium (or $^3He$). There are some minor differences: the total four momentum of the interacting nucleons con have a transverse component. 

\subsection{Two different target nucleons interact with large transverse momentum exchange}
\subsubsection {General features and diagonal terms}

As already seen, when two or more target nucleons interact with large transverse momentum exchange, nuclear and partonic dynamics are interconnected. The presence of the nuclear wave function induces in fact the presence of two kinds of contributions: a "diagonal term" and a number of nondiagonal or "interference" terms.
 The diagonal discontinuity for the triple scattering is shown in fig. 6
       and its analytical expression is: 
       
 \begin{eqnarray}
 Disc \;{\cal A}^{(2,1)}\big|_d &=&
{1\over {(2\pi)^{35}}}\int {{\check\phi_p}\over{l_1^2 l_2^2l_3^2}}
{{{\check\phi_p}^*}\over{{l'}_1^2 {l'}_2^2{l'}_3^2}}
{{\hat\phi_1}\over{a_1^2 a_2^2 }}{{\phi_2}\over{ a_3^2}}
{{{\hat\phi_1}^*}\over{{a'}_1^2 {a'}_2^2}} {{\phi_2}^*\over{ a_3^2}} \cr
 &\times&T_1(l_1a_1\to q_1 q'_1) T_2(l_2a_2\to q_2q'_2) T_3(l_3a_3\to q_3q'_3)\cr
 &\times&{T_1}^*({l'}_1{a'}_1\to q_1q'_1) T_2^*(l_2a_2\to q_2q'_2){T_3}^*({l'}_3{a'}_3\to q_3q'_3)\cr
 &\times&{{\Phi_D(N_1,N_2)}\over{[N_1^2-m^2][N_2^2-m^2]}}
{{{\Phi_D}^*({N'}_1,{N'}_2)}\over{[{N'}_1^2-m^2][{N'}_2^2-m^2]}}\cr
&\times& \delta(L-\sum l-F_4) \delta(N_1-a_3-F_1) \delta(N_2-a_1-a_2-F_2)\cr
&\times&\delta(L-\sum {l'}-F_4)\delta({N'}_1-{a'}_3-F_1)\delta ({N'}_2-{a'}_1-{a'}_2-F_2)\cr
&\times&\delta(l_1+a_1-Q_1)\delta({l'}_1+{a'}_1-Q_1)\delta(l_2+a_2-Q_2)
\delta({l'}_2+{a'}_2-Q_2)\cr
&\times&\delta(l_3+a_3-Q_3)\delta({l'}_3+{a'}_3-Q_3)
\delta(D-N_1-N_2)\delta(D-{N'}_1-{N'}_2)\cr
&\times& \prod d(\Omega/8) dad{a'}dld{l'}dFdN d{N'} dQ_i
 \end{eqnarray}

\noindent
with the mass shell condition $D^2=M_D^2$.\par

In the case of a Tritium or $^3He$ (with four-momentum $T$)  the corresponding discontinuities $ Disc\;{\cal B}^{(2,1,0)}$ are obtained from $Disc \;{\cal A}^{(2,1)}$ by substituting the factor

\begin{eqnarray}
 {{\Phi_D(N_1,N_2)}\over{[N_1^2-m^2][N_2^2-m^2]}}
{{{\Phi_D}^*({N'}_1,{N'}_2)}\over{[{N'}_1^2-m^2][{N'}_2^2-m^2]}}\delta(D-N_1-N_2)\delta(D-{N'}_1-{N'}_2)
\end{eqnarray}

\noindent
with

\begin{eqnarray}
{{\Phi_{T}(N_1,N_2,N_3)}\over{[N_1^2-m^2][N_2^2-m^2][N_3^2-m^2]}}&&
{{{\Phi_{T}}^*({N'}_1,{N'}_2,N_3)}\over{[{N'}_1^2-m^2][{N'}_2^2-m^2][{N'}_3^2-m^2]}}\\
\times&&\delta(T-N_1-N_2-N_3)\delta(T-{N'}_1-{N'}_2-N_3)\delta (N_3^2-m^2)\nonumber
\end{eqnarray}

\noindent
The integration runs over $dN_i$, with $i=1,2,3$, and the mass-shell condition is: $ T^2=M_{T}^2$.
\par
 From the side of the incoming proton the situation and the subsequent manipulations are the same as in the previous case, they give rise to the factor $\psi_3$. From the side of the nucleus we find different structures: the one-body and two-body parton vertices and the singularities which put a nucleon on mass shell or the partons on mass shell.

\begin{figure}[h]
\centering
\includegraphics[width=160mm]{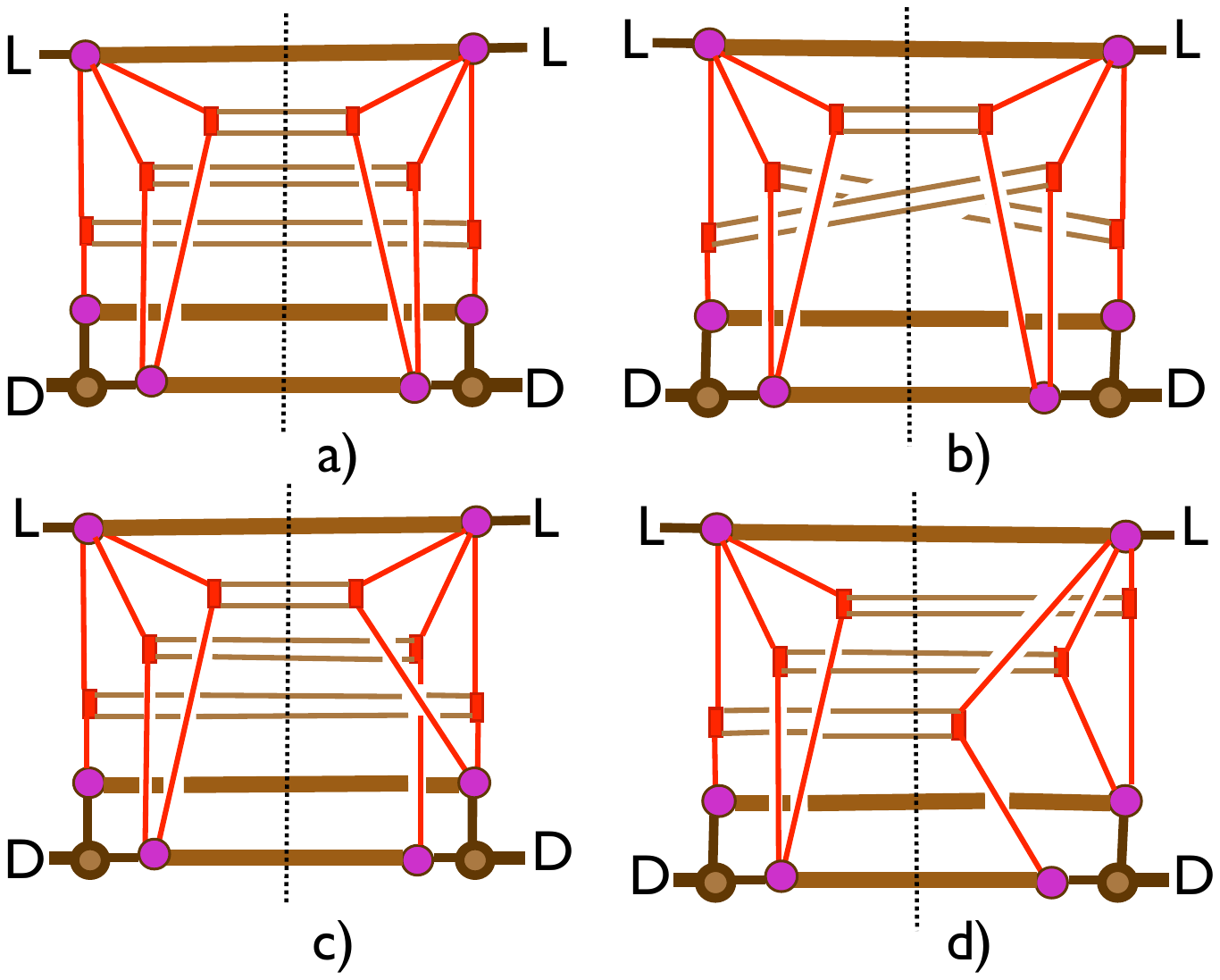}
\caption {Different contributions to the triple parton scattering in $p\ D$ interactions. Both target nucleons interact with large transverse momentum exchange} \label{210}
\end{figure}

Here we need the one-parton and the two-parton wave functions, they were already defined and an explicit form was given in eq, (5). The longitudinal
integration of the remnants is performed as in the previous case. We have now three $F_{\pm}$ and we get:
 $$1/{F_1}_-=1/[N_-(1-z_1-z_2)]\qquad 1/{F_2}_-=1/[N_- (1-z_3)]\qquad 1/{F_4}_+=1/[L_+(1-\sum x)]\;.$$

 Concerning the nuclear variables: since the binding energy is small, the most important singularities are those corresponding to the nucleons' mass shell condition.
 For the Deuteron one must thus evaluate $1/(N_1^2-m^2)$ with the other propagator on mass shell, $i.e.\, N_2^2=m^2 $, with $N_1+N_2=D$. One obtains
 the expression of $\Psi_D(N_-)/N_-$ as given in Eq (6).
 For the Tritium or $^3He$ we have $N_1+N_2+N_3=T$ and, in addition, the mass-shell condition for the spectator, we get the expression in eq. (7).
 The one-two-three parton densities together with their Fourier transform in the transverse plain $\Gamma(x_i;b_i)$ have been already defined and discussed.\par
Neglecting terms of order $1/\sqrt s$, the conservation of the large components gives again ${l_i}_+={{l'}_i}_+,\;{a_1}_-={{a'}_1}_-$ and the complete expression is brought into a diagonal form by Fourier transforming the transverse variables. In this case, the geometrical relations are however different. One finds: $\beta_2-\beta_3=b_2-b_3-B_1+B_2$, $\beta_1-\beta_2=b_1-b_2+B_1-B_2$. The corresponding configuration in transverse space is shown in Fig 7a).
 \par
The diagonal contribution to the triple parton scattering cross sections on Deuteron and on Tritium or $^3He$, when one nucleon interacts twice and another once, setting
$B=B_1-B_2$ and $B'=B_1-B_3$, are thus expressed as:

\begin{eqnarray}
\sigma^{pD}_{3,2}\big|_{d}&=&{2\over{(2\pi)^3}} \int\Gamma(x_1,x_2,x_3;\beta_1,\beta_2,\beta_3)\Gamma(z_1,z_2;b_1,b_2)
\Gamma(z_3;b_3){{d\hat\sigma}\over{d\Omega_1}}{{d\hat\sigma}\over{d\Omega_2}}{{d\hat\sigma}\over{d\Omega_3}}\cr
&&|\Psi_D(Z;B)|^2dB
d\beta_1d\beta_2d\beta_3\,db_1db_2db_3\,\delta(b_1-b_2+\beta_1-\beta_2)\,\delta(b_1-b_3-\beta_1+\beta_3-B)\cr
&&dx_1dx_2dx_3\,dz_1dz_2dz_3\,d\Omega_1d\Omega_2d\Omega_3\, 
dZ/Z^2\;.\cr
\cr
\sigma^{pT}_{3,2}\big|_{d}&=&{6\over{(2\pi)^3}}\int\Gamma(x_1,x_2,x_3;\beta_1,\beta_2,\beta_3)
\Gamma(z_1,z_2;b_1,b_2)
\Gamma(z_3;b_3){{d\hat\sigma}\over{d\Omega_1}}{{d\hat\sigma}\over{d\Omega_2}}{{d\hat\sigma}\over{d\Omega_3}}\\
&& |\Psi_T(Z_i;B,B')|^2dB dB'
d\beta_1d\beta_2d\beta_3\,db_1db_2db_3\cr
&&\delta(b_1-b_2+\beta_1-\beta_2)\,\delta(b_1-b_3-\beta_1+\beta_3-B)\;
 dx_1dx_2dx_3\,dz_1dz_2dz_3\cr && d\Omega_1d\Omega_2d\Omega_3\,
\delta(Z_1+Z_2+Z_3-3)dZ_1
dZ_2 dZ_3/(Z_1^2Z_2)\;.\nonumber
\end{eqnarray}

\subsubsection{Interference terms}

 As it appears from the graphs in Fig.6, there are three kind of interference terms which differ one another for the different relations between the partons and the parent nucleon.

 \par
This difference takes, therefore, the form of difference in the $\delta-$functions: the discontinuities corresponding to three terms are explicitly given for the case of the deuteron: they can be summarized in the following form:

 \begin{eqnarray}
 Disc \;{\cal A}^{(2,1)}\big|_{in} &=&
{1\over {(2\pi)^{35}}}\int {{\check\phi_p}\over{l_1^2 l_2^2l_3^2}}
{{{\check\phi_p}^*}\over{{l'}_1^2 {l'}_2^2{l'}_3^2}}
{{\hat\phi_1}\over{a_1^2 a_2^2 }}{{\phi_2}\over{ a_3^2}}
{{\phi_1}^*\over{ {a'}_1^2}}{{{\hat\phi_2}^*}\over{{a'}_2^2 {a'}_3^2}}  \cr
 &\times&T_1(l_1a_1\to q_1 q'_1) T_2(l_2a_2\to q_2q'_2) T_3(l_3a_3\to q_3q'_3)\cr
 &\times&{T_1}^*({l'}_1{a'}_1\to q_1q'_1) T_2^*(l_2a_2\to q_2q'_2){T_3}^*({l'}_3{a'}_3\to q_3q'_3)\cr
 &\times&{{\Phi_D(N_1,N_2)}\over{[N_1^2-m^2][N_2^2-m^2]}}
{{{\Phi_D}^*({N'}_1,{N'}_2)}\over{[{N'}_1^2-m^2][{N'}_2^2-m^2]}}\cr
&\times& \delta(L-\sum l-F_4)\delta(L-\sum {l'}-F_4)\times \eta(N_i,a_j) \cr
&\times&\delta(l_1+a_1-Q_1)\delta({l'}_1+{a'}_1-Q_1)\delta(l_2+a_2-Q_2)
\delta({l'}_2+{a'}_2-Q_2)\cr
&\times&\delta(l_3+a_3-Q_3)\delta({l'}_3+{a'}_3-Q_3)
\delta(D-N_1-N_2)\delta(D-{N'}_1-{N'}_2)\cr
&\times& \prod d\Omega dad{a'}dld{l'}dFdN d{N'} dQ
\end{eqnarray}

The three relevant realizations of the factor $\eta$, as it can be seen from the graphs, are:

\begin{eqnarray}
\eta_{2,1}(N_i,a_j)&=&\delta (N_1-a_3-F_1)\delta(N_2-a_1-a_2-F_2)\delta (N'_1-a'_2-F_1)\delta (N'_2-a'_1-a'_3-F_2)\qquad\cr
\eta_{2,2}(N_i,a_j)&=&\delta (N_1-a_3-F_1)\delta(N_2-a_1-a_2-F_2)\delta (N'_1-a'_1-a'_3-F_1)\delta (N'_2-a'_2-F_2)\qquad\cr
\eta_{2,3}(N_i,a_j)&=&\delta (N_1-a_3-F_1)\delta(N_2-a_1-a_2-F_2)\delta (N'_1-a'_1-a'_2-F_1)\delta (N'_2-a'_3-F_2)\qquad
\end{eqnarray}

 The corresponding expression for the case of the Tritium are obtained by the same substitutions that were used in the diagonal term.  
 \par
 The configuration produced by the factors $\eta_{2,j}$ are similar to the configuration described by the crossed diagram in the double scattering, however the factors are more strictly interlocked so that it is necesssary to introduce other auxiliary terms

\begin{eqnarray}
 &W_{2,1}&(\bar x_1,\bar x_2,\bar x_3;Z_1,Z_2;b_1,b_2,b_3;B)=
 \frac{1}{4(2\pi)^9}\frac{\bar x_1\bar x_2\bar x_3}{(Z_1-\bar x_1)(Z_2-\bar x_2- \bar x_3)}\cr
 &\times&\int\tilde\psi_{1, M_1}(\bar x_1/Z_1;b_1)
 \tilde\psi_{2, M_2}(\bar x_2/Z_2,\bar x_3/Z_2;b_2,b_3)\cr
 &\times&\tilde\psi_{2, M_2}^*(\bar x_1/Z'_2,\bar x_3/Z'_2;b_3,b_2+B) \tilde\psi_{1, M_1}^*(\bar x_2/Z'_1;B-b_1)
 dM_1^2\,dM_2^2
 \end{eqnarray}

 The nuclear factors have formally the same expression as in the diagonal case,
 but the values of $Z$ are different on the two sides, while keeping the constraints $Z_1+Z_2=Z'_1+Z'_2=2$ . We have in fact $Z_1-\bar x_1=Z'_1-\bar x_2,\quad Z_2-\bar x_2=Z'_2-\bar x_1$.\par

\begin{figure}[h]
\centering
\includegraphics[width=160mm]{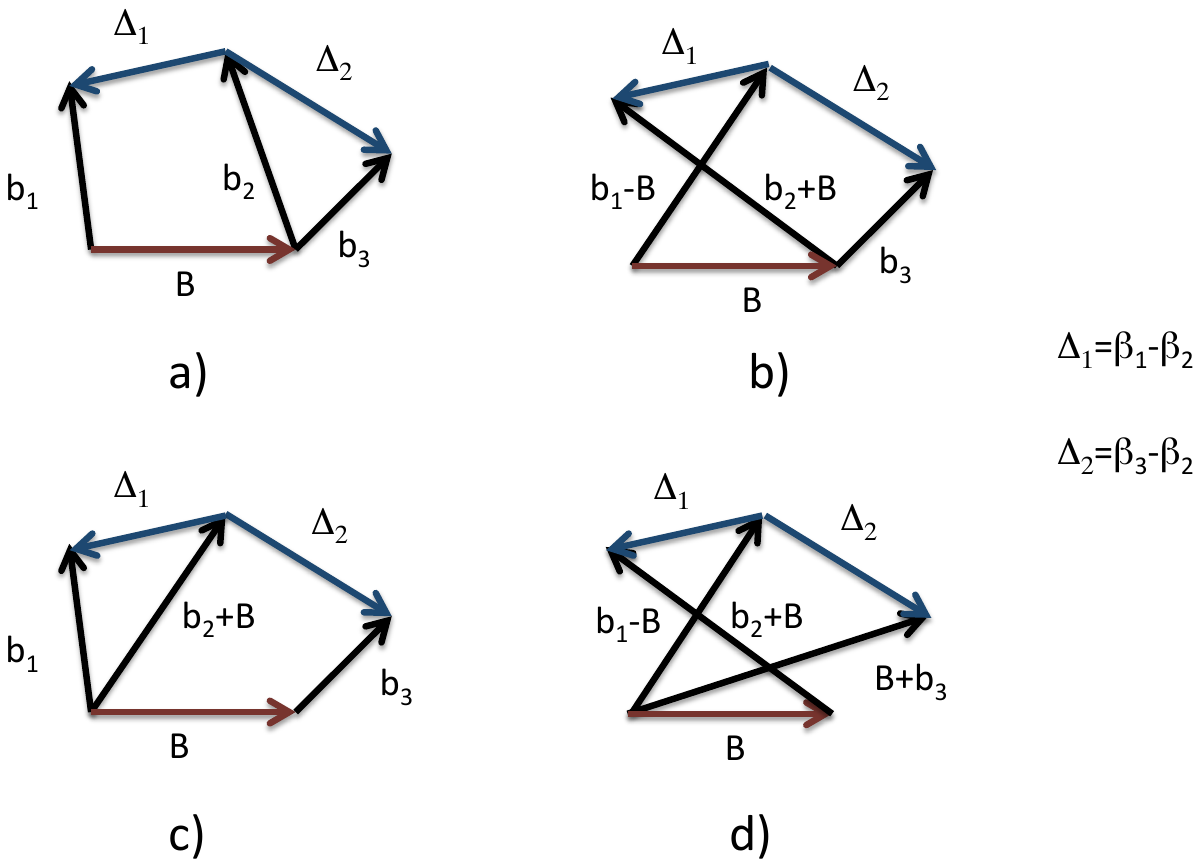}
\caption {Configurations in transverse space of the four amplitudes in Fig.6, in the right hand side of the cut} \label{Fig32}
\end{figure}

\begin{eqnarray}
 &W_{2,2}&(\bar x_1,\bar x_2,\bar x_3;Z_1,Z_2;b_1,b_2,b_3;B)=
 \frac{1}{4(2\pi)^9}\frac{\bar x_1\bar x_2\bar x_3}{(Z_1-\bar x_1)(Z_2-\bar x_2- \bar x_3)}\cr
 &\times&\int\tilde\psi_{1, M_1}(\bar x_1/Z_1;b_1)\tilde\psi_{2, M_2}(\bar x_2/Z_2,
\bar x_3/Z_2;b_2,b_3)\cr
&\times&\tilde\psi_{2, M_1}^*(\bar x_1/Z'_1,\bar x_3/Z'_1;b_2+B,b_1) \tilde\psi_{1, M_2}^*(\bar x_2/Z'_2;b_3)
 dM_1^2\,dM_2^2
 \end{eqnarray}

 Here beyond the the constraints $Z_1+Z_2=Z'_1+Z'_2=2$ we have the relations
 $Z_1=Z'_1-\bar x_3,\quad Z_2=Z'_2+\bar x_3$.\par

  \begin{eqnarray}
 &W_{2,3}&(\bar x_1,\bar x_2,\bar x_3;Z_1,Z_2;b_1,b_2,b_3;B)=
 \frac{1}{4(2\pi)^9}\frac{\bar x_1\bar x_2\bar x_3}{(Z_1-\bar x_1)(Z_2-\bar x_2- \bar x_3)}\cr
 &\times&\int\tilde\psi_{1, M_1}(\bar x_1/Z_1;b_1)\tilde\psi_{2, M_2}(\bar x_2/Z_2,\bar x_3/Z_2;b_3)\cr
 &\times&\tilde\psi_{2, M_1}^*(\bar x_2/Z'_1,\bar x_3/Z'_1;b_2+B,b_3+B) \tilde\psi_{1, M_1}^*(\bar x_1/Z'_2;B-b_1) dM_1^2\,dM_2^2
 \end{eqnarray}

 Here beyond the the constraints $Z_1+Z_2=Z'_1+Z'_2=2$ we have the relations
 $Z_1-\bar x_1=Z'_1-\bar x_2-\bar x_3,\quad Z_2-\bar x_2-\bar x_3=Z'_2-\bar x_1$.

 \par
 We saw that the nuclear factors have formally the same expression as in the diagonal part, but the values of $Z$ are different on the two sides. This feature was already found in the cross diagram for double scattering, and so also the qualitative consideration are of the same kind. Precisely the longitudinal variables are on one side $Z_1$ and $Z_2$ whereas on the other side they are $Z'_1$ and $Z'_2$. In the nonrelativistic conditions of the internal motion, in particular
 $N_{\bot}^2<<m^2$, which are the actual conditions in the Deuteron and Tritium the typical width of the nuclear wave function of the Deuteron, in dimensionless variables, is of the order $\sqrt{(4m^2-M_D^2)/(M_D^2)}$, while the differences $Z-Z'$ are of the order of the fractional momenta $\bar x$.\par
 Now we see that the cross sections for these particular processes can be obtained from the expression of the diagonal term (eq.25) by substituting the factors $\Gamma(z_1,z_2;b_1,b_2)\Gamma(z_3;b_3)$ with the correspnding $W_{2,j}$ term, after solving the constraints which give $Z'_1, Z'_2$ in terms of $Z_1, Z_2$.\par
In comparing the diagonal term with the interference terms, one can repeat the same qualitative considerations made for the double scattering case, $i.e.$ in the partonic amplitudes the denominators $Z,Z'$ could be set equal to 1, while their value is relevant in the nuclear functions; the transverse variables are of the order of the hard size of the hadron, so they are relatively small as compared with the nuclear size.

\subsection{Three different target nucleons interact with large transverse momentum exchange}
\subsubsection {General features and diagonal terms}

 This kind of process can evidently happen only with a three-body nucleus (at least), here also the presence of the nuclear wave function induces the presence of two kinds of contributions: a "diagonal term" and a number of nondiagonal or "interference" terms.

 \begin{figure}[h]
\centering
\includegraphics[width=160mm]{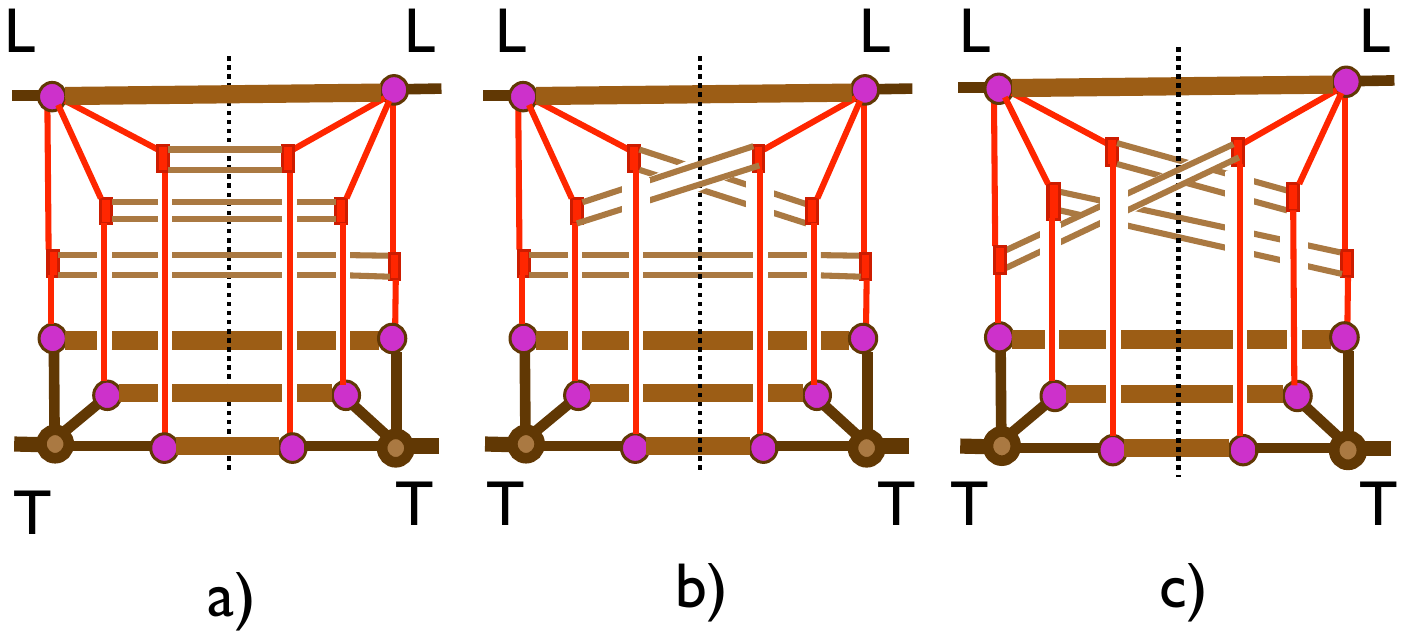}
\caption {Different contributions to triple parton scattering in $p\ T$ or $p\ ^3He$ interactions. All target nucleons interact with large transverse momentum exchange} \label{111}
\end{figure}

\begin{figure}[h]
\centering
\includegraphics[width=160mm]{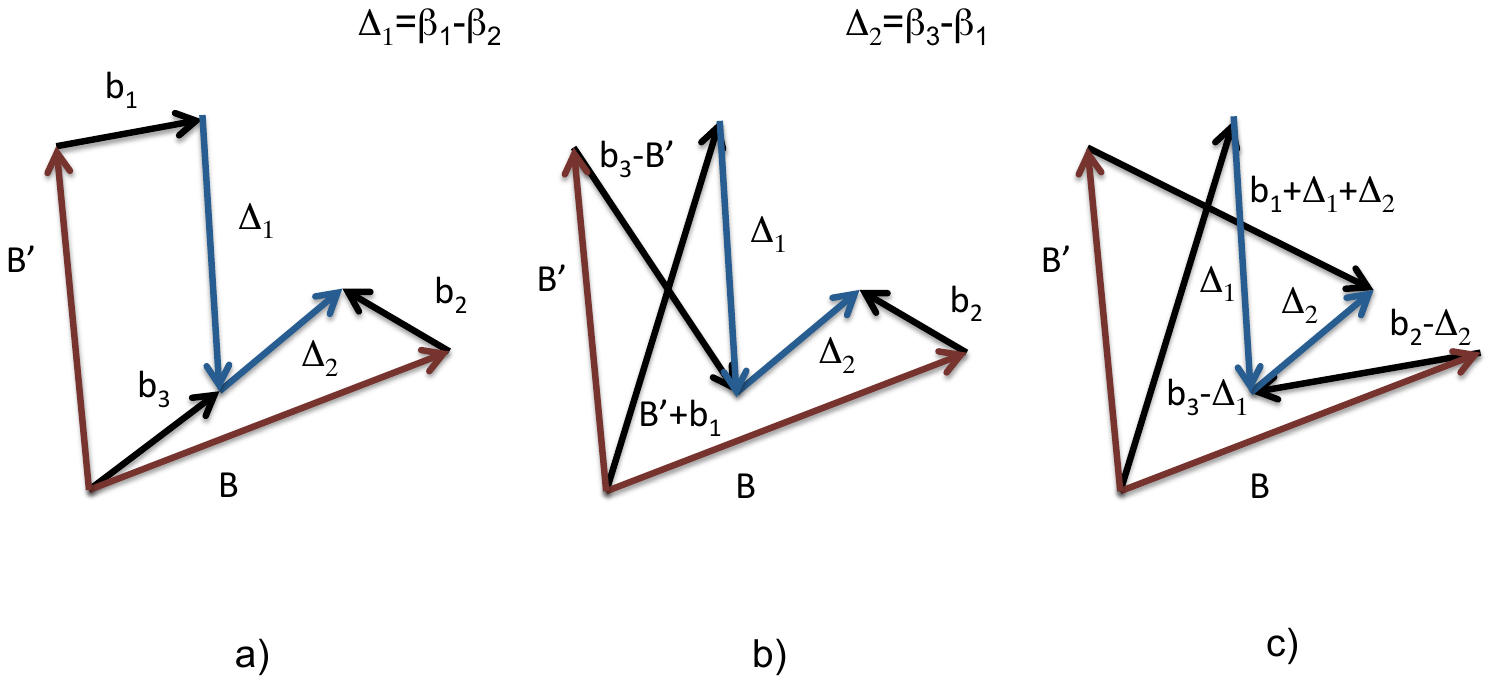}
\caption {Configurations in transverse space of the three amplitudes in Fig.8, in the right hand side of the cut} \label{Fig33}
\end{figure}

 The diagonal discontinuity for the triple scattering is:

 \begin{eqnarray}
 Disc&\;&{\cal B}^{(1,1,1)}\big|_d=
{1\over {(2\pi)^{38}}}\int {{\check\phi_p}\over{l_1^2 l_2^2l_3^2}}
{{{\check\phi_p}^*}\over{{l'}_1^2 {l'}_2^2{l'}_3^2}}
{{\phi_1}\over{a_1^2 }}{{\phi_2}\over{ a_2^2}}{{\phi_3}\over{ a_3^2}}
{{{\hat\phi_1}^*}\over{{a'}_1^2 {a'}_2^2}} {{\phi_3}^*\over{ a_3^2}} \cr
 &\times&T_1(l_1a_1\to q_1 q'_1) T_2(l_2a_2\to q_2q'_2) T_3(l_3a_3\to q_3q'_3)\cr
 &\times&{T_1}^*({l'}_1{a'}_1\to q_1q'_1) T_2^*(l_2a_2\to q_2q'_2){T_3}^*({l'}_3{a'}_3\to q_3q'_3)\cr
 &\times&{{\Phi_D(N_1,N_2,N_3)}\over{[N_1^2-m^2][N_2^2-m^2][N_3^2-m^2]}}
{{{\Phi_D}^*({N'}_1,{N'}_2),{N'}_3)}\over{[{N'}_1^2-m^2][{N'}_2^2-m^2][{N'}_3^2-m^2]}}\cr
&\times& \delta(L-\sum l-F_4) \delta(N_1-a_1-F_1) \delta(N_2-a_2-F_2)\delta(N_3-a_3-F_3)\cr
&\times&\delta(L-\sum {l'}-F_4)\delta({N'}_1-{a'}_1-F_1)\delta ({N'}_2-{a'}_2-F_2)\delta ({N'}_3-{a'}_3-F_3)\cr
&\times&\delta(l_1+a_1-Q_1)\delta({l'}_1+{a'}_1-Q_1)\delta(l_2+a_2-Q_2)
\delta({l'}_2+{a'}_2-Q_2)\cr
&\times&\delta(l_3+a_3-Q_3)\delta({l'}_3+{a'}_3-Q_3)
\delta(T-N_1-N_2-N_3)\delta(T-{N'}_1-{N'}_2-{N'}_3)\cr
&\times& \prod d(\Omega/8) dad{a'}dld{l'}dFdN d{N'} dQ
 \end{eqnarray}
\par 

\noindent
and the mass shell condition is $T^2=M_T^2$.\par
From the projectile side we have the three-parton densities, which have been already defined and used. On the Tritium side we must use the full three-body structure of the nuclear wave function:
 the conservation $N_i=a_i+F_i$ gives $a_i^2=(N_i-F_i)_+{a_i}_--{a_i}_{\bot}^2$ then  the integral to be performed may be written as:
 $${\cal Y}_a= \int d{N_1}_+d{N_2}_+d{N_3}_+\delta(\sum_{i=1,2,3}{N_i}_+-T_+)\prod_{i=1,2,3}\frac{1}{{{N_i}_+{N_i}_- {m_{\bot}}_i}^2}{1\over {{(N-F)_i}_+{a_i}_- -{a_i}^2_{\bot}}} $$

\noindent
  As in the previous cases there are two kinds of singularities in the integrand, one kind puts a nucleon on mass shell, the other one puts on mass shell the parton, and here also the singularities putting the nucleons on mass shell are the most important ones. So we approximate ${\cal Y}_a$ as:

 \begin{equation}
{\cal Y}_a\approx -(2\pi)^2{1\over{T_+-I_1-I_2-I_3}}
 \prod_{i=1,2,3}\frac{1}{[(m_{\bot}^2-N_-F_+)_i {a_i}_--(N_-a_{\bot}^2)_i]}\qquad
 I_j={m_{\bot}}_j^2/{N_-}_j  \end{equation}

When considering the complete cut graph we find that the relation between $l,a,Q$ yields again the equalities $l_+={l'}_+,\;a_-={a'}_-$, whereas the equality of the remnants yields $N_-={N'}_-$, these can be converted into the fractional momenta $Z$, with the constraint $ Z_1+ Z_2+ Z_3=3$.
Here also it is convenient to go from transverse momenta to transverse coordinates.

\par
 The factors $\psi$ and $\Gamma$
 are defined as before.
Since in ${\cal Y}_a$ the dependences on $a_{\bot}$ and on $N_{\bot}$ are interlocked, the Fourier transform must be performed with respect to both sets of transverse variables $a_{\bot}$ and $N_{\bot}$.

\par
 As before $\beta_i$ are conjugated to $l_i$, $b_i$ are conjugated to $a_i$  and $B_i$ are conjugated to $N_i$.
 The conservations, that will be again conveniently expressed in exponential form, are:
 $l_i+a_i={l'}_i+{a'}_i$, $\sum l_i=\sum {l'}_i$, $N_i+a_i={N'}_1+{a'}_i$,
 $\sum N_i=\sum {N'}_i=T$.
 The integrations over the transverse momenta yield the equalities $\beta_i=\beta'_i$, $b_i=b'_i$, $B_i=B'_i$ and the geometrical conditions $b_i+B_i-\beta_i=const$ that can be expressed also as: $b_i+B_i-\beta_i=b_j+B_j-\beta_j;\;i\neq j$ \par
We can write, for the diagonal term:

\begin{eqnarray}
\sigma^{pT}_{3,3}\big|_{d}&=&{3\over{(2\pi)^3}}\int\Gamma(x_1,x_2,x_3;\beta_1,\beta_2,\beta_3)
\Gamma(z_1;b_1)\Gamma(z_2;b_2)
\Gamma(z_3;b_3){{d\hat\sigma}\over{d\Omega_1}}{{d\hat\sigma}\over{d\Omega_2}}{{d\hat\sigma}\over{d\Omega_3}}\cr
&& |\Psi_T(Z_i;B,\underline{ B})|^2dB d\underline{ B}
d\beta_1d\beta_2d\beta_3\,db_1db_2db_3\cr
&& \delta(b_1-b_3-\beta_1+\beta_3+\underline{ B})\,\delta(b_1-b_2-\beta_1+\beta_2+B)\;
 dx_1dx_2dx_3\,dz_1dz_2dz_3\cr && d\Omega_1d\Omega_2d\Omega_3\,
\delta(Z_1+Z_2+Z_3-3)dZ_1
dZ_2 dZ_3/(Z_1Z_2)^2\;.
\label{eq:S33}
\end{eqnarray}

In the vertex function only the differences $B_i-B_j$ are relevant, therefore the integration variables $B,\underline{ B}$ represent a pair of these differences, $e.g.\; B=B_1-B_2,\; \underline{ B}=B_1-B_3$.\par

 \subsubsection {Interference terms}
 The discontinuity for the interference terms in triple scattering can be
 written, as in the previous chapter:

 \begin{eqnarray}
 Disc&\;&{\cal B}^{(1,1,1)}\big|_i=
{1\over {(2\pi)^{38}}}\int {{\check\phi_p}\over{l_1^2 l_2^2l_3^2}}
{{{\check\phi_p}^*}\over{{l'}_1^2 {l'}_2^2{l'}_3^2}}
{{\phi_1}\over{a_1^2 }}{{\phi_2}\over{ a_2^2}}{{\phi_3}\over{ a_3^2}}
{{{\hat\phi_1}^*}\over{{a'}_1^2 {a'}_2^2}} {{\phi_3}^*\over{ a_3^2}} \cr
 &\times&T_1(l_1a_1\to q_1 q'_1) T_2(l_2a_2\to q_2q'_2) T_3(l_3a_3\to q_3q'_3)\cr
 &\times&{T_1}^*({l'}_1{a'}_1\to q_1q'_1) T_2^*(l_2a_2\to q_2q'_2){T_3}^*({l'}_3{a'}_3\to q_3q'_3)\cr
 &\times&{{\Phi_D(N_1,N_2,N_3)}\over{[N_1^2-m^2][N_2^2-m^2][N_3^2-m^2]}}
{{{\Phi_D}^*({N'}_1,{N'}_2),{N'}_3)}\over{[{N'}_1^2-m^2][{N'}_2^2-m^2][{N'}_3^2-m^2]}}\cr
&\times& \delta(L-\sum l-F_4) \delta(L-\sum {l'}-F_4) \times \eta(N_i,a_i)\cr
&\times&\delta(l_1+a_1-Q_1)\delta({l'}_1+{a'}_1-Q_1)\delta(l_2+a_2-Q_2)
\delta({l'}_2+{a'}_2-Q_2)\cr
&\times&\delta(l_3+a_3-Q_3)\delta({l'}_3+{a'}_3-Q_3)
\delta(T-N_1-N_2-N_3)\delta(T-{N'}_1-{N'}_2-{N'}_3)\cr
&\times& \prod d(\Omega/8) dad{a'}dld{l'}dFdN d{N'} dQ_i
 \end{eqnarray}

 Here we find two essentially different realizations of the factors $\eta$

 \begin{eqnarray}
 \eta_{3,1}(N_i,a_i)&=&\delta(N_1-a_1-F_1)\delta(N_2-a_2-F_2)\delta(N_3-a_3-F_3)\cr
           &&\delta(N'_1-a'_3-F_1)\delta(N'_2-a'_2-F_2)\delta(N'_3-a'_1-F_3)\cr
  \eta_{3,2}(N_i,a_i)&=&\delta(N_1-a_1-F_1)\delta(N_2-a_2-F_2)\delta(N_3-a_3-F_3)\cr
           &&\delta(N'_1-a'_2-F_1)\delta(N'_2-a'_3-F_2)\delta(N'_3-a'_1-F_3)
\end{eqnarray}

\noindent
and this leads to the definition of two more auxiliary functions $W$.

\begin{eqnarray}
W_{3,1}&&(\bar x_1,\bar x_2,\bar x_3;Z_1,Z_2,Z_3;b_1,b_2,b_3;B,\underline{ B})
=\frac{1}{4(2\pi)^9}\frac{\bar x_1\bar x_2\bar x_3}{(Z_1-\bar x_1)(Z_2-\bar x_2- \bar x_3)}\cr
\times&&\tilde\psi_{1, M_1}(\bar x_1/Z_1;b_1)\tilde\psi_{1, M_2}(\bar x_2/Z_2;b_2)\tilde\psi_{1, M_3}(\bar x_3/Z_3;b_3)\cr
\times&&\tilde\psi^*_{1, M_3}(\bar x_1/Z_1;b_1+\underline{ B})\tilde\psi^*_{1, M_2}(\bar x_2/Z_2;b_2)\tilde\psi^*_{1, M_1}(\bar x_3/Z_3;b_3-\underline{ B})
\end{eqnarray}

 Here beyond the the constraints $Z_1+Z_2+Z_3=Z'_1+Z'_2+Z'_3=3$ we have the relations
 $Z_1-\bar x_1=Z'_1-\bar x_3,\; Z_2=Z'_2,\; Z_3-\bar x_3=Z'_3-\bar x_1$.
 \par
\begin{eqnarray}
W_{3,2}&&(\bar x_1,\bar x_2,\bar x_3;Z_1,Z_2,Z_3;b_1,b_2,b_3;B,\underline{ B})=
 \frac{1}{4(2\pi)^9}\frac{\bar x_1\bar x_2\bar x_3}{(Z_1-\bar x_1)(Z_2-\bar x_2- \bar x_3)}\cr
\times&&\tilde\psi_{1, M_1}(\bar x_1/Z_1;b_1)\tilde\psi_{1, M_2}(\bar x_2/Z_2;b_2)\tilde\psi_{1, M_3}(\bar x_3/Z_3;b_3)\cr
\times&&\tilde\psi^*_{1, M_3}(\bar x_1/Z_1;b_3-\Delta_1)\tilde\psi^*_{1, M_2}(\bar x_2/Z_2;b_2-\Delta_2)
\tilde\psi^*_{1, M_1}(\bar x_3/Z_3;b_1+\Delta_2+\Delta_1)
\end{eqnarray}

 Here beyond the the constraints $Z_1+Z_2+Z_3=Z'_1+Z'_2+Z'_3=3$ we have the relations
 $Z_1-\bar x_1=Z'_1-\bar x_2,\; Z_2-\bar x_2=Z'_2-\bar x_3\; Z_3-\bar x_3=Z'_3-\bar x_1$.\par
 Finally the cross sections for these particular processes can be obtained from the expression of the diagonal term, Eq.\eqref{eq:S33}, by substituting the factors $\Gamma(z_1;b_1)\Gamma(z_2;b_2)\Gamma(z_3;b_3)$ with the correspnding $W_{3,j}$ term, after solving the constraints which give $Z'_1, Z'_2, Z'_3$ in terms of $Z_1, Z_2, Z_3$. The transverse configurations are shown in Fig.9.

 \par
 Clearly the qualitative considerations made previously for ratio between the diagonal and the interference term hold also here, since they depend on the existence of two scales (hadronic and nuclear) playing always the same role.

\section {Simplest estimates of the dominant contributions}

The non-perturbative component of MPI in $pA$ collisions is characterized by the hadronic and by the nuclear scale. In MPI, the relevant hadronic scale is the transverse dimension $R$ of the generalized parton distributions, which is smaller as compared with the hadron radius and it may be roughly a factor four smaller as compared with the radii of $D$, $^3H$ and $^3He$. Even with light nuclei, one may thus obtain a simplest estimate of the dominant contributions to the MPI cross sections by neglecting the hadronic scale when compared to the nuclear scale. In the same spirit one may obtain a further simplification by evaluating the integrals on the fractional momenta of the bound nucleons, $Z_i$, by keeping into account the dependence on $Z_i$ only in the nuclear wave function and replacing $Z_i$ with $1$ anywhere else.

\subsection{Double scattering}

The double parton scattering cross section, for $p\ D$ and of $p\ ^3H$ (or $p\ ^3He$) collisions, are given by the sum of two contributions, where only a single bound nucleon and where two bound nucleons participate in the hard interaction:

\begin{eqnarray}
\sigma^{pD}_{2}(x_i,\bar x_i)&=&\sigma^{pD}_{2,1}(x_i,\bar x_i)+\sigma^{pD}_{2,2}(x_i,\bar x_i)\cr
\sigma^{pT}_{2}(x_i,\bar x_i)&=&\sigma^{pT}_{2,1}(x_i,\bar x_i)+\sigma^{pT}_{2,2}(x_i,\bar x_i)
\end{eqnarray}

With the simplifying assumptions above, the contributions where only a single bound nucleon participates are given by

\begin{eqnarray}
\sigma^{pD}_{2,1}\simeq 2\  \sigma_D\qquad{\rm and}\qquad\sigma^{pT}_{2,1}\simeq 3\  \sigma_D
\end{eqnarray}

\noindent
where $\sigma_D$ is the double parton scattering inclusive cross section on a isolated nucleon.
One can make the positions:

\begin{eqnarray}
\Gamma(x_1,x_2;\beta_1,\beta_2)&\equiv&K_{x_1,x_2}G(x_1)G(x_2)f_{x_1,x_2}(\beta_1,\beta_2)\cr
{\rm with}&&\int f_{x_1,x_2}(\beta_1,\beta_2)d\beta_1d\beta_2=1
\end{eqnarray}

\noindent
where $G(x)$ is the one body inclusive parton distribution, such that $G(x)=\int\Gamma(x;b)db$. In the case of identical interactions, one may thus express the double parton scattering cross section on a isolated nucleon as

\begin{eqnarray}
\sigma_D(x_1,\bar x_1, x_2,\bar x_2)&=&\frac{1}{2}K_{x_1,x_2}K_{\bar x_1,\bar x_2}\int f_{x_1,x_2}(\beta_1,\beta_2)f_{\bar x_1,\bar x_2}(b_1,b_2)\sigma_S(x_1,\bar x_1)\sigma_S(x_2,\bar x_2)\cr
&&\qquad\qquad\qquad\quad\times\ \delta(\beta_1-\beta_2-b_1+b_2)d\beta_1d\beta_2db_1db_2
\end{eqnarray}

\noindent
where $\sigma_S=\int G(x)\hat\sigma(x,x')G(x')dxdx'$ is the single parton scattering inclusive cross section on a isolated nucleon. The effective cross section, namely the accessible experimental information in nucleon-nucleon collisions, is thus given by

\begin{eqnarray}
\frac{1}{\sigma_{eff}(x_1,\bar x_1, x_2,\bar x_2)}=K_{x_1,x_2}K_{\bar x_1,\bar x_2}\int f_{x_1,x_2}(\beta_1,\beta_1-\Delta_1)f_{\bar x_1,\bar x_2}(b_1,b_1-\Delta_1)d\beta_1db_1d\Delta_1
\label{eq:Sig.eff}
\end{eqnarray}

\noindent 
where $\Delta_1=\beta_1-\beta_2$. In $p\ D$ collisions, when two nucleons participate to the hard interaction, one has contributions from a diagonal and from a off diagonal term. The
dominant contribution to the diagonal term is

\begin{eqnarray}
\sigma^{pD}_{2,2}\big|_d(x_1,\bar x_1,x_2,\bar x_2)&=&\frac{1}{(2\pi)^3}\int\Gamma(x_1,x_2;\beta_1,\beta_2)
\hat\sigma(x_1,\bar x_1)
\hat\sigma(x_2,\bar x_2)
\Gamma(\bar x_1/ Z;b_1)\Gamma(\bar x_2/(2- Z);b_2)\cr
&\times &|\tilde\Psi_D( Z;B)|^2 dZ/Z^2\ dB\,db_1\, db_2\,d\beta_1\,d\beta_2\delta(B-b_1 +b_2-\Delta_1)\cr
&\simeq& K_{x_1,x_2}\sigma_S(x_1,\bar x_1)\sigma_S(x_2,\bar x_2){\cal I}_D(0)
\label{eq:SD2,2}
\end{eqnarray}

\noindent
where the general form of ${\cal I}_D(x)$ is:

\begin{eqnarray}
{\cal I}_D(x)&\equiv&\frac{1}{(2\pi)^3}\int\tilde\Psi_D( Z;0) \tilde\Psi_D^*( Z';0)\frac{dZdZ'}{ZZ'}\delta(Z-Z'-x)
\end{eqnarray}

The range of the variables $b_i$ and $\beta_j$, defined by the partonic distributions $\Gamma$, is much narrower as compared with the nuclear range of $B$. The $\delta$-function in Eq.\eqref{eq:SD2,2} thus forces, in $\tilde\Psi_D$, $B\approx 0$.

Notice that, differently from the case of double parton scattering in nucleon-nucleon  collisions,  the scale factor  in $\sigma^{pD}_{2,2}\big|_d$ is given by the value of ${\cal I}_D(0)$, which is  determined by the radius of the Deuteron, while the cross section is proportional to $K_{x_1,x_2}$, which gives the partonic correlation in fractional momenta. As already noticed in \cite{Calucci:2010wg}, $\sigma^{pD}_{2,2}\big|_d$ depends thus weakly on the correlation between partons in the transverse coordinates and, on the contrary, it may provide a rather direct information on the size of  $K_{x_1,x_2}$.

The contribution of the interference term to the cross section is

\begin{eqnarray}
\sigma^{pD}_{2,2}\big|_i(x_1,\bar x_1,x_2,\bar x_2)&=&\frac{1}{(2\pi)^3}\int\Gamma(x_1,x_2;\beta_1,\beta_2)
\hat\sigma(x_1,\bar x_1)
\hat\sigma(x_2,\bar x_2)
W_1(Z,Z';\bar x_1,\bar x_2;b_1,b_2,B)\cr
& \times&\tilde\Psi_D( Z;B) \tilde\Psi_D^*( Z';B)\big[ZZ'\big]^{-1}dB\,\,db_1\, db_2\,d\beta_1\, d\beta_2 dZdZ'\cr
&\times&\delta(B-b_1 +b_2-\Delta_1)\delta(Z-Z'-\bar x_1+\bar x_2)
\label{eq:S22I}
\end{eqnarray}

By neglecting the hadron scale with respect to the nuclear scale and keeping $Z\not=1$ only in the Deuteron wave function, the integrations in $b_1$ and $b_2$ in Eq.\eqref{eq:S22I} are

\begin{eqnarray}
 && \int W_1(1,1;\bar x_1,\bar x_2;b_1,b_2,b_1 -b_2+\Delta_1)db_1\ db_2\cr
 &&\qquad=\frac {1}{4(2\pi)^6} \int dM_1^2dM_2^2 db_1\ db_2
 \frac {\bar x_1\bar x_2}{(1-\bar x_1)(1-\bar x_2)}\psi_{M_1}(\bar x_1;b_1) \psi^*_{M_2}(\bar x_2;b_2-\Delta_1)\cr
 &&\qquad\qquad\qquad\qquad\qquad\qquad\qquad\qquad\qquad\qquad\qquad\times\psi_{M_2}(\bar x_1;b_2) \psi^*_{M_1}(\bar x_2;b_1+\Delta_1)\cr
 &&\qquad=\tilde H(\bar x_1,\bar x_2;\Delta_1)\tilde H(\bar x_2,\bar x_1;-\Delta_1)
 \label{eq:W1}
  \end{eqnarray}

\noindent
where the generalized parton distributions $\tilde H$ have been introduced:

\begin{eqnarray}
\tilde H(\bar x_1,\bar x_2;\Delta_1)\equiv\int H(1,1;\bar x_1,\bar x_2;b_1,b_2)\delta(b_1-b_2-\Delta_1)db_1\ db_2
 \label{eq:H}
 \end{eqnarray}

Notice that the normalization is $\tilde H( x, x; 0)=G(x)$. One may thus define

\begin{eqnarray}
C_1(x_1,\bar x_1, x_2,\bar x_2)=\frac{\int f_{x_1,x_2}(\Delta_1)\tilde H(\bar x_1,\bar x_2;\Delta_1)\tilde H(\bar x_2,\bar x_1;-\Delta_1)\ d\Delta_1}{G(\bar x_1)G(\bar x_2)}
\label{eq:fhh}
\end{eqnarray}

\noindent
which is dimensionless and weakly dependent of $\bar x_1,\ \bar x_2$ as compared to ${\cal I}_D(\bar x_1- \bar x_2)$, since $C_1$ originates from the partonic structure of the hadron, while ${\cal I}_D$ originates from the nuclear structure. The contribution of the interference term to the cross section may thus be expressed as

\begin{equation}
\sigma^{pD}_{2,2}\big|_i(x_1,\bar x_1,x_2,\bar x_2)
\simeq K_{x_1,x_2}\sigma_S(x_1,\bar x_1)\sigma_S(x_2,\bar x_2)\ C_1(x_1,\bar x_1,x_2,\bar x_2){\cal I}_D(\bar x_1-\bar x_2)
\end{equation}

Notice that both $\sigma^{pD}_{2,2}\big|_d$ and $\sigma^{pD}_{2,2}\big|_i$ depend linearly on $K_{x_1,x_2}$ and both terms are proportional to the inverse of the square of the Deuteron radius, the latter term through the nuclear off diagonal factor ${\cal I}_D(\bar x_1-\bar x_2)$, which induces a much stronger dependence of  $\sigma^{pD}_{2,2}\big|_i$  on $\bar x_1-\bar x_2$ as compared with $\sigma^{pD}_{2,2}\big|_d$.

In the case of double parton interactions in $p\ ^3H$ or $p\ ^3He$ collisions, with two target nucleons taking part to the hard interaction, one obtains the same expressions, for the dominant contributions to the cross sections, as in the case of $p\ D$ interactions. The only difference is in the multiplicity factors and in the terms ${\cal I}_D(0)$ and ${\cal I}_D(\bar x_1-\bar x_2)$, which are replaced by the corresponding quantities with $^3H$ or $^3He$, actually ${\cal I}_T(0)$ and ${\cal I}_T(\bar x_1-\bar x_2)$.

The leading contributions to the double parton scattering cross sections in $p\ D$ and $p\ ^3H$, $p\ ^3He$ are thus given by

\begin{eqnarray}
\sigma^{pD}_{2}(x_i,\bar x_i)&=&\sigma^{pD}_{2,1}(x_i,\bar x_i)+\sigma^{pD}_{2,2}|_d(x_i,\bar x_i)+\sigma^{pD}_{2,2}|_i(x_i,\bar x_i)\cr
&\simeq& 2\  \sigma_D(x_i,\bar x_i)+K_{x_1,x_2}\sigma_S(x_1,\bar x_1)\sigma_S(x_2,\bar x_2)
\bigl[{\cal I}_D(0)+C_1(x_i,\bar x_i){\cal I}_D(\bar x_1-\bar x_2)\bigr]\cr
\sigma^{pT}_{2}(x_i,\bar x_i)&=&\sigma^{pT}_{2,1}(x_i,\bar x_i)+\sigma^{pT}_{2,2}|_d(x_i,\bar x_i)+\sigma^{pT}_{2,2}|_i(x_i,\bar x_i)\\
&\simeq& 3\  \sigma_D(x_i,\bar x_i)+3\ K_{x_1,x_2}\sigma_S(x_1,\bar x_1)\sigma_S(x_2,\bar x_2)
\bigl[{\cal I}_T(0)+C_1(x_i,\bar x_i){\cal I}_T(\bar x_1-\bar x_2)\bigr]\nonumber
\end{eqnarray}

The contributions $\sigma^{pD}_{2,1}$ and $\sigma^{pT}_{2,1}$ are well approximated by $2\  \sigma_D$ and by $3\ \sigma_D$. The actual values can be evaluated with great accuracy, once the double parton scattering cross sections in $pp$ and in $pn$ collisions are known as a function of fractional momenta. Also the off diagonal nuclear terms  ${\cal I}_D(0)$, ${\cal I}_D(\bar x_1-\bar x_2)$, ${\cal I}_T(0)$ and ${\cal I}_T(\bar x_1-\bar x_2)$ can be evaluated very accurately. By measuring the double parton scattering cross sections in $p\ D$ and $p\ ^3H$ (or $p\ ^3He$) one may thus obtain accurate estimates of the  differences $\sigma^{pD}_{2}-\sigma^{pD}_{2,1}$ and  $\sigma^{pT}_{2}-\sigma^{pT}_{2,1}$ and, as a consequence, of the rations ${\cal R}_D$ and ${\cal R}_T$, defined as

\begin{eqnarray}
{\cal R}_D(x_i,\bar x_i)\equiv\frac{\sigma^{pD}_{2}(x_i,\bar x_i)-\sigma^{pD}_{2,1}(x_i,\bar x_i)}{\sigma_S(x_1,\bar x_1)\sigma_S(x_2,\bar x_2)}&\simeq&K_{x_1,x_2}
\bigl[{\cal I}_D(0)+C_1(x_i,\bar x_i){\cal I}_D(\bar x_1-\bar x_2)\bigr]\cr
{\cal R}_T(x_i,\bar x_i)\equiv\frac{\sigma^{pT}_{2}(x_i,\bar x_i)-\sigma^{pT}_{2,1}(x_i,\bar x_i)}{\sigma_S(x_1,\bar x_1)\sigma_S(x_2,\bar x_2)}&\simeq&3\ K_{x_1,x_2}
\bigl[{\cal I}_T(0)+C_1(x_i,\bar x_i){\cal I}_T(\bar x_1-\bar x_2)\bigr]
\label{eq:R}
\end{eqnarray}

One should point out that the contribution of the interference term is not always present. As an example, in the case of production of a $W+$jets, through double parton collisions, the interference term is absent. In such a case, the second term in square brackets in ~\eqref{eq:R} is missing and Eq.~\eqref{eq:R} allows a direct estimate of $K_{x_1,x_2}$, namely of the importance of the correlations in $x$ in the double parton distributions. When the interference term is present, the cross section depends on the additional unknown quantity, $C_1(x_i,\bar x_i)$, which multiplies the nuclear overlap integrals ${\cal I}_D(\bar x_1-\bar x_2)$ or ${\cal I}_T(\bar x_1-\bar x_2)$, which originate the main dependence of the cross section on $X\equiv\bar x_1-\bar x_2$. By measuring the cross section at different values of $X$ one may construct the fraction

\begin{eqnarray}
\frac{\Delta{\cal R}_{D,T}(x_i,\bar x_i)}{\Delta X}&\simeq&K_{x_1,x_2}C_1(x_i,\bar x_i)\frac{\Delta{\cal I}_{D,T}(X)}{\Delta X}
\label{eq:R1}
\end{eqnarray}

By studying the dependence of ${\cal R}_D$ and of ${\cal R}_T$ on $X=\bar x_1-\bar x_2$, with the help of Eq.s~\eqref{eq:R} and \eqref{eq:R1}, one may obtain information both on $C_1(x_i,\bar x_i)$ and on $K_{x_1,x_2}$. Notice that the values of $\Delta X$ needed in Eq.\eqref{eq:R1} may not be too small. The natural scale of $X=\bar x_1-\bar x_2$ is in fact $(E_B/m)^{1/2}\approx5\times 10^{-2} $, where $E_B$ is the nuclear binding energy.

The indication on the value of $K_{x_1,x_2}$, together with the measure of the effective cross section in nucleon-nucleon collisions, allow to obtain an indication on the value of the integral $\int f_{x_1,x_2}(\Delta_1)f_{\bar x_1,\bar x_2}(\Delta_1)d\Delta_1$ (cfr Eq.~\eqref{eq:Sig.eff}). With the help of Eq.s\eqref{eq:R} and \eqref{eq:R1}, one may estimate $C_1(x_i,\bar x_i)$ and obtain an indication on the value of the integral $\int f_{x_1,x_2}(\Delta_1)\tilde H(\bar x_1,\bar x_2;\Delta_1)\tilde H(\bar x_2,\bar x_1;-\Delta_1)\ d\Delta_1$ (cfr Eq.~\eqref{eq:fhh}). The information on the two integrals will provide important constraints on the correlation length between partons in transverse space, which is explicit in $f_{x_1,x_2}(\Delta_1)$. By measuring the double parton scattering cross section in $p\ D$ and $p\ ^3H$ (or $p\ ^3He$) one may thus learn both on correlations between partons in fractional momenta, through the factor $K_{x_1,x_2}$, and on correlations between partons in the transverse coordinates.

\subsection{Triple scattering}

As in the case of double parton scattering, the triple parton scattering cross section in $p\ D$ and of $p\ ^3H$ or $p\ ^3He$ collisions can be written as

\begin{eqnarray}
\sigma^{pD}_{3}(x_i,\bar x_i)&=&\sigma^{pD}_{3,1}(x_i,\bar x_i)+\sigma^{pD}_{3,2}(x_i,\bar x_i)\cr
\sigma^{pT}_{3}(x_i,\bar x_i)&=&\sigma^{pT}_{3,1}(x_i,\bar x_i)+\sigma^{pT}_{3,2}(x_i,\bar x_i)+\sigma^{pT}_{3,3}(x_i,\bar x_i)
\end{eqnarray}

With the simplifying assumptions discussed in the previous chapter, the dominant contribution to the terms where only a single bound nucleon undergoes a triple parton interaction is given by

\begin{eqnarray}
\sigma^{pD}_{3,1}\simeq 2\  \sigma_T,\qquad\sigma^{pT}_{3,1}\simeq 3\  \sigma_T
\end{eqnarray}

\noindent
where $\sigma_T$ is the triple parton scattering inclusive cross section on a isolated nucleon.
By making the positions:

\begin{eqnarray}
\Gamma(x_1,x_2,x_3;\beta_1,\beta_2,\beta_3)\equiv
K_{x_1,x_2,x_3}G(x_1)G(x_2)G(x_3)f_{x_1,x_2,x_3}(\beta_1,\beta_2,\beta_3)
\end{eqnarray}

\noindent
with

\begin{eqnarray}
\int f_{x_1,x_2,x_3}(\beta_1,\beta_2,\beta_3)d\beta_1d\beta_2d\beta_3=1
\end{eqnarray}

\noindent
one may express, in the case of identical interactions, the triple parton scattering cross section on a isolated nucleon as

\begin{eqnarray}
\sigma_T(x_1,\bar x_1, x_2,\bar x_2, x_3,\bar x_3)&=&\frac{1}{6}K_{x_1,x_2,x_3}K_{\bar x_1,\bar x_2,\bar x_3}\int f_{x_1,x_2,x_3}(\beta_i)f_{\bar x_1,\bar x_2,\bar x_3}(b_i)\\
&\times&\delta(\beta_1-\beta_2-b_1+b_2)\delta(\beta_1-\beta_3-b_1+b_3)\prod \sigma_S(x_i,\bar x_i)d\beta_idb_i
\nonumber
\end{eqnarray}

\subsubsection{Two different target nucleons interact with large transverse momentum exchange}

The contribution to the triple parton scattering cross section, where two target nucleons undergo hard interactions, is a process of ${\cal O}\bigl(1/(S^2R^2)\bigr)$, to be compared with triple scattering on a single nuceon, which is of ${\cal O}\bigl(1/R^4\bigr)$, $R$ and $S$ are the hadronic and nuclear scales. In the case of $p\ D$ collisions, the different contributions to the cross section are summarized by the expression:

\begin{eqnarray}
\sigma^{pD}_{3,2}\big|_{j}&=&{2\over{(2\pi)^3}} {\cal N}_j\int\Gamma(x_1,x_2,x_3;\beta_1,\beta_2,\beta_3)
{{d\hat\sigma}\over{d\Omega_1}}{{d\hat\sigma}\over{d\Omega_2}}{{d\hat\sigma}\over{d\Omega_3}}\cr
&&\qquad\qquad\times\ W_{2,j}(\bar x_1,\bar x_2,\bar x_3;Z_1,Z_2;b_1,b_2,b_3;B)
\tilde\Psi_D( Z_1,Z_2;B) \tilde\Psi_D^*( Z_1',Z_2';B)\cr
&&\qquad\qquad\times\ dB\,\,dZ_1dZ_1'dZ_2dZ_2'/(Z_1Z_1')\prod_idx_id\bar x_idb_i\,d\beta_i d\Omega_idZ_i\cr
&&\qquad\qquad\times\ \delta(Z_1+Z_2-2)\delta(Z_1'+Z_2'-2)\delta(Z_1-Z_1'-X_j)\delta(Z_2-Z_2'+X_j)\cr
&&\qquad\qquad\times\ \delta(b_2-b_1+\Delta_1)\delta(B-b_1+b_3-\Delta_2)
\label{eq:S3,2}
\end{eqnarray}

\noindent
where $\Delta_1=\beta_1-\beta_2$ and $\Delta_2=\beta_3-\beta_1$, while the index $j$ corresponds to the diagonal case, when $j=0$, and to the the three different interference terms, when $j=1,2,3$. By neglecting the hadronic scale $R$ as compared to the nuclear scale $S$, in the diagonal case one obtains

\begin{eqnarray}
W_{2,0}(\bar x_1,\bar x_2,\bar x_3;Z_1,Z_2;b_1,b_2,b_3;B)\equiv\frac{1}{2}\Gamma(z_1,z_2;b_1,b_2)\Gamma(z_3;b_3)
\end{eqnarray}

The quantity $X_j$ assumes the following values:

\begin{eqnarray}
X_0=0,\qquad X_1=\bar x_1-\bar x_2,\qquad X_2=-\bar x_2,\qquad X_3=\bar x_1-\bar x_2-\bar x_3
\end{eqnarray}

\noindent
and the multiplicity factors ${\cal N}_j$ are: ${\cal N}_0=2,\ {\cal N}_1=4,\ {\cal N}_2=4,\ {\cal N}_3=2$.

Analogously to the case previously discussed, the dominant contributions may be estimated by

\begin{eqnarray}
\sigma^{pD}_{3,2}\big|_{j}&\simeq&{\cal N}_j\int\Gamma(x_1,x_2,x_3;\beta_1,\beta_2,\beta_3)
\hat\sigma(x_1,\bar x_1)\hat\sigma(x_2,\bar x_2)\hat\sigma(x_3,\bar x_3)\cr
&&\qquad\qquad\times\ W_{2,j}(\bar x_1,\bar x_2,\bar x_3;1,1;b_1,b_2,b_3;b_1-b_3+\Delta_2)
{\cal I}_D(X_j)\cr
&&\qquad\qquad\times\ \delta(b_2-b_1+\Delta_1)\prod_idx_id\bar x_idb_i\,d\beta_i
\end{eqnarray}

\noindent
where all effects of the Deuteron wave function are summarized in the terms ${\cal I}_D(X_j)$:

\begin{eqnarray}
{\cal I}_D(X_j)={1\over{(2\pi)^3}} \int
\tilde\Psi_D( Z;0) \tilde\Psi_D^*( Z';0)\delta(Z-Z'-X_j)\frac{dZdZ'}{ZZ'}
\end{eqnarray}

\noindent
already defined in Eq.45.
By introducing

\begin{eqnarray}
C_{2,j}(x_i,\bar x_i)=\frac{\int f_{x_1,x_2,x_3}(\beta_i)W_{2,j}(\bar x_i;1,1;b_i;b_1-b_3+\Delta_2)\delta(b_2-b_1+\Delta_1)\prod_i db_i\beta_i}{G(\bar x_1)G(\bar x_2)G(\bar x_3)}
\end{eqnarray}

\noindent
one obtains

\begin{eqnarray}
\sigma^{pD}_{3,2}\big|_{j}(x_i,\bar x_i)&\simeq&{\cal N}_jK_{x_i}\sigma_S(x_1,\bar x_1)\sigma_S(x_2,\bar x_2)\sigma_S(x_3,\bar x_3)C_{2,j}(x_i,\bar x_i){\cal I}_D(X_j)
\end{eqnarray}

The expression of the diagonal contribution is

\begin{eqnarray}
\sigma^{pD}_{3,2}\big|_{0}(x_i,\bar x_i)&\simeq&\frac{1}{2}K_{x_1,x_2,x_3}K_{\bar x_1\bar x_2}\sigma_S(x_1,\bar x_1)\sigma_S(x_2,\bar x_2)\sigma_S(x_3,\bar x_3){\cal G}_{x_i\bar x_k}{\cal I}_D(0)
\label{eq:overlap32}
\end{eqnarray}

\noindent
where

\begin{eqnarray}
{\cal G}_{x_i\bar x_k}&=&\int f_{x_1,x_2,x_3}(\beta_1,\beta_2,\beta_3)f_{\bar x_1\bar x_2}(b_1,b_2)\delta(b_2-b_1+\Delta_1)\prod d\beta_idb_k
\label{eq:d32}
\end{eqnarray}

As in the case of double collisions one may introduce the ratio

\begin{eqnarray}
{\cal R}_D'(x_i,\bar x_i)&\equiv&\frac{\sigma^{pD}_{3,2}(x_i,\bar x_i)}{\prod\sigma_S(x_i,\bar x_i)}=
\frac{\sum_j\sigma^{pD}_{3,2}\big|_j(x_i,\bar x_i)}{\prod\sigma_S(x_i,\bar x_i)}\simeq K_{x_i}\sum_j{\cal N}_jC_{2,j}(x_i,\bar x_i){\cal I}_D(X_j)
\label{eq:R1D}
\end{eqnarray}

\noindent
and, for $j>0$, construct the fraction

\begin{eqnarray}
\frac{\Delta{\cal R}_{D}'(x_i,\bar x_i)}{\Delta X_j}&\simeq&K_{x_i}{\cal N}_jC_{2,j}(x_i,\bar x_i)\frac{\Delta{\cal I}_{D}'(X_j)}{\Delta X_j}
\label{eq:R1D1}
\end{eqnarray}

Differently with respect to the case of the double collisions, Eq.s\eqref{eq:R1D} and \eqref{eq:R1D1} do not allow to disentangle $K_{x_i}$ from $C_{2,j}(x_i,\bar x_i)$. Disentangling the effects of longitudinal and transverse correlations was possible in the case of double collisions because, in that case, the dominant contribution to the diagonal term depends only on $K_{x_1,x_2}$. In the actual case, on the contrary, $C_{2,0}(x_i,\bar x_i)$ is proportional to the product $K_{x_i}{\cal G}_{x_i\bar x_k}$. By studying triple scattering on Deuteron one may only obtain an estimate of the products $K_{x_i}C_{2,j}(x_i,\bar x_i)$. To gain further insight into longitudinal and transverse three-body correlations one needs additional information, which can be provided by triple parton interactions in collisions of protons with $^3H$ or with $^3He$.

When two target nucleons participate to the hard interaction, after integrating on the spectator nucleon, $p\ ^3H$ (or $p\ ^3He$) give very similar results as $p\ D$ collisions. The two dominant contributions to the diagonal term differ in fact only by an overall multiplicity factor (which is actually 3) and in the factors ${\cal I}_T(X_j)$, which substitute the factors ${\cal I}_D(X_j)$. Analogously to the case of $D$, ${\cal I}_T(X_j)$ represent the square of the $^3H$ (or $^3He$) wave function in the mixed representation, integrated in the fractional momenta $Z_i$ with the constraints given in Eq.\eqref{eq:S3,2}, and in the relative transverse distance $B'$, while the transverse distance $B$ has been set equal to zero. One thus obtains the relation

\begin{eqnarray}
\sigma^{pT}_{3,2}\big|_{j}(x_i,\bar x_i)\simeq\sigma^{pD}_{3,2}\big|_{j}(x_i,\bar x_i)\frac{{\cal I}_T(X_j)}{{\cal I}_D(X_j)}
\label{eq:S3,21}
\end{eqnarray}

Eq.\eqref{eq:S3,21} is a consequence of Eq.\eqref{eq:R1D}, which holds in the limit $R^2/S_D^2\to 0$. If $S_T$ is the $^3H$ (or $^3He$) radius, the two terms in Eq.\eqref{eq:S3,21} are thus of ${\cal O}\Bigl(1/(R^2S_T^2)\Bigr)$ and the relation is exact in the limit $R^2/S_D^2\to 0$. Finite values of $R$ contribute, in the left hand side, with terms of ${\cal O}\Bigl(1/(R^2S_T^2)\times R^2/S_T^2\Bigr)$ and with terms of ${\cal O}\Bigl(1/(R^2S_T^2)\times R^2/S_D^2\Bigr)$ in the right hand side of the equation. One may thus estimate that Eq.\eqref{eq:S3,21} is valid up to terms of ${\cal O}\Bigl(1/S_T^2\times (1/S_T^2-1/S_D^2)\approx(S_D^2-S_T^2)/S_T^6\Bigr)$.

\subsubsection{Three different target nucleons interact with large transverse momentum exchange}

In the case of the contribution to the triple parton scattering cross section, $\sigma^{pT}_{3,3}$, where three different target nucleons interact with large transverse momentum exchange, in $p\ ^3H$ or $p\ ^3He$ collisions one has three different terms, one diagonal and two off-diagonal, which are labeled with the index $j$ in the expression here below. As in the previous section, the label $j=0$ corresponds to the diagonal case.

\begin{eqnarray}
\sigma^{pT}_{3,3}\big|_{j}&=&{2\over{(2\pi)^3}} {\cal N}_j'\int\Gamma(x_1,x_2,x_3;\beta_1,\beta_2,\beta_3)
{{d\hat\sigma}\over{d\Omega_1}}{{d\hat\sigma}\over{d\Omega_2}}{{d\hat\sigma}\over{d\Omega_3}}\cr
&&\qquad\qquad\times\ W_{3,j}(\bar x_1,\bar x_2,\bar x_3;Z_1,Z_2, Z_3;b_1,b_2,b_3;B,B')
\tilde\Psi_T( Z_1,Z_2,Z_3;B, B')\cr
&&\qquad\qquad\times\  \tilde\Psi_T^*( Z_1,'Z_2',Z_3';B,B')
dB\,\,dZ_1dZ_1'dZ_2dZ_2'dZ_3dZ_3'/(Z_1Z_1'Z_2Z_2')\cr
&&\qquad\qquad\times\ \delta(Z_1+Z_2+Z_3-3)\delta(Z_1'+Z_2'+Z_3'-3)\delta(Z_1-Z_1'-Y_{j,1})\delta(Z_2-Z_2'+Y_{j,2})\cr
&&\qquad\qquad\times\ \delta(Z_3-Z_3'+Y_{j,3})\ \delta(B'+b_1-b_3+\Delta_1)\,\delta(b_3-b_2-\Delta_2-B)\cr
&&\qquad\qquad\times\ \prod_idx_id\bar x_idb_i\,d\beta_i d\Omega_idBdB'
\end{eqnarray}

With the approximations previously discussed one obtains

\begin{eqnarray}
\sigma^{pT}_{3,3}\big|_{j}&\simeq&{2\over{(2\pi)^3}} {\cal N}_j'\int\Gamma(x_1,x_2,x_3;\beta_1,\beta_2,\beta_3)
\hat\sigma(x_1,\bar x_1)\hat\sigma(x_2,\bar x_2)\hat\sigma(x_3,\bar x_3)\cr
&&\qquad\qquad\times\ W_{3,j}(\bar x_1,\bar x_2,\bar x_3;1,1, 1;b_1,b_2,b_3;b_3-b_2-\Delta_2,\ -b_1+b_3-\Delta_1)\cr
&&\qquad\qquad\times{\cal J}(Y_{j, i=1,3})\prod_idx_id\bar x_idb_i\,d\beta_i
\end{eqnarray}

\noindent
where

\begin{eqnarray}
{\cal J}(Y_{j, i=1,3})&=&{2\over{(2\pi)^3}} {\cal N}_j'\int
\tilde\Psi_T( Z_1,Z_2,Z_3;0, 0) \tilde\Psi_T^*( Z_1',Z_2',Z_3';0,0)\cr
&&\qquad\qquad\times\ dZ_1dZ_1'dZ_2dZ_2'dZ_3dZ_3'\big[Z_1Z_1'Z_2Z_2'Z_3Z_3'\big]^{-1/2}\cr
&&\qquad\qquad\times\ \delta(Z_1+Z_2+Z_3-3)\delta(Z_1'+Z_2'+Z_3'-3)\cr
&&\qquad\qquad\times\ \delta(Z_1-Z_1'-Y_{j,1})\delta(Z_2-Z_2'+Y_{j,2})
\end{eqnarray}

\noindent
and

\begin{center}
  \begin{tabular}{ | l | l  l  l | }
    \hline
    $Y_{j=0,i=1,3}\ :\ $ & $Y_{0,1}=0\ ;$ & $Y_{0,2}=0\ ;$  & $Y_{0,3}=0$  \\ \hline
    $Y_{j=1,i=1,3}\ :\ $ & $Y_{1,1}=\bar x_1-\bar x_3\ ;$ & $Y_{1,2}=0\ ;$  & $Y_{1,3}=\bar x_3-\bar x_1$  \\ \hline
    $Y_{j=2,i=1,3}\ :\ $ & $Y_{2,1}=\bar x_1-\bar x_2\ ;$ & $Y_{2,2}=\bar x_2-\bar x_3\ ;$  & $Y_{2,3}=\bar x_3-\bar x_1$ \\
    \hline
  \end{tabular}
\end{center}

\noindent
while the multiplicity factors are ${\cal N}_0'=1$, ${\cal N}_1'=3$, ${\cal N}_1'=2$. Introducing

\begin{eqnarray}
&&C_{3,j}(x_i,\bar x_i)=\cr
&&\frac{\int f_{x_1,x_2,x_3}(\beta_i)
W_{3,j}(\bar x_i;1,1, 1;b_i;b_3-b_2-\Delta_2,\ -b_1+b_3-\Delta_1)\prod_idb_i\,d\beta_i}{G(\bar x_1)G(\bar x_2)G(\bar x_3)}
\end{eqnarray}

\noindent
one obtains

\begin{eqnarray}
\sigma^{pT}_{3,3}(x_i,\bar x_i)\big|_{j}&\simeq&{\cal N}_j'K_{x_1,x_2,x_3}
\sigma_S(x_1,\bar x_1)\sigma_S(x_2,\bar x_2)\sigma_S(x_3,\bar x_3)C_{3,j}(x_i,\bar x_i)
{\cal J}(Y_{j, i=1,3})
\end{eqnarray}

In the case $j=0$ one has

\begin{eqnarray}
C_{3,0}(x_i,\bar x_i)=1
\end{eqnarray}

\noindent
and one obtains

\begin{eqnarray}
\sigma^{pT}_{3,3}\big|_{0}(x_i,\bar x_i)&\simeq&\frac{1}{6}K_{x_1,x_2,x_3}
\sigma_S(x_1,\bar x_1)\sigma_S(x_2,\bar x_2)\sigma_S(x_3,\bar x_3)
{\cal J}(0)
\label{eq:S33}
\end{eqnarray}

For $j=1$ one has

\begin{eqnarray}
&&\int W_{3,1}(\bar x_i;1,1,1;b_i;b_3-b_2-\Delta_2,-b_1+b_3-\Delta_1)\prod db_i\cr
&&\qquad\qquad\qquad\qquad\qquad\qquad=\tilde H(\bar x_1,\bar x_3;\Delta_1)G(x_2)\tilde H(\bar x_3,\bar x_1;-\Delta_1)
\end{eqnarray}

\noindent
in such a way that

\begin{eqnarray}
C_{3,1}(x_i,\bar x_i)=\frac{\int f_{x_1,x_2,x_3}(\Delta_1)\tilde H(\bar x_1,\bar x_3;\Delta_1)\tilde H(\bar x_3,\bar x_1;-\Delta_1) d\Delta_1}{G(\bar x_1)G(\bar x_3)}
\end{eqnarray}

\noindent
where

\begin{eqnarray}
f_{x_1,x_2,x_3}(\Delta_1)=\int f_{x_1,x_2,x_3}(\beta_i) \delta(\Delta_1-\beta_1+\beta_2)\prod d\beta_i
\end{eqnarray}

For $j=2$ one has

\begin{eqnarray}
&&\int W_{3,2}(\bar x_i;1,1,1;b_i;b_3-b_2-\Delta_2,-b_1+b_3-\Delta_1)\prod db_i\cr
&&\qquad\qquad\qquad\qquad\qquad\qquad=\tilde H(\bar x_1,\bar x_3;\Delta_1+\Delta_2)\tilde H(\bar x_2,\bar x_1;-\Delta_2)\tilde H(\bar x_3,\bar x_2;-\Delta_1)
\end{eqnarray}

\noindent
and

\begin{eqnarray}
&&C_{3,2}(x_i,\bar x_i)=\cr
&&\qquad\frac{\int f_{x_1,x_2,x_3}(\Delta_1,\Delta_2)\tilde H(\bar x_1,\bar x_3;\Delta_1+\Delta_2)\tilde H(\bar x_2,\bar x_1;-\Delta_2)\tilde H(\bar x_3,\bar x_2;-\Delta_1) d\Delta_1d\Delta_2}{G(\bar x_1)G(\bar x_2)G(\bar x_3)}
\end{eqnarray}

\noindent
where

\begin{eqnarray}
f_{x_1,x_2,x_3}(\Delta_1,\Delta_2)=\int f_{x_1,x_2,x_3}(\beta_i) \delta(\Delta_1-\beta_1+\beta_2)\delta(\Delta_2+\beta_1-\beta_3)\prod d\beta_i
\end{eqnarray}

As in the case of double parton collisions, discussed previously, the contribution to the triple parton scattering cross section $\sigma^{pT}_{3,1}$ is given with good approximation by $3\ \sigma_T$, where $\sigma_T$ is the triple parton scattering cross section on a isolated nucleon. Once the triple parton scattering cross sections in $pp$ and in $pn$ collisions are known as a function of fractional momenta, the smearing effects of the nuclear wave function can be taken into account and $\sigma^{pT}_{3,1}$ can be evaluated with great accuracy. Also the off diagonal nuclear terms  ${\cal J}(Y_{j, i=1,3})$ can be evaluated with great accuracy. By measuring $\sigma^{pT}_{3}$, one may thus obtain an accurate estimate of the  difference $\sigma^{pT}_{3}-\sigma^{pT}_{3,1}$. Eq.\eqref{eq:S3,21} allows to estimate $\sigma^{pT}_{3,2}$. One may thus define

\begin{eqnarray}
{\cal R}_T'(x_i,\bar x_i)&\equiv&
\frac{\sigma^{pT}_{3,3}(x_i,\bar x_i)}{\prod\sigma_S(x_i,\bar x_i)}
\end{eqnarray}

\noindent
where

\begin{eqnarray}
\sigma^{pT}_{3}(x_i,\bar x_i)&=&\sigma^{pT}_{3,1}(x_i,\bar x_i)+\sigma^{pT}_{3,2}(x_i,\bar x_i)+\sigma^{pT}_{3,3}(x_i,\bar x_i)\cr
\sigma^{pT}_{3,2}(x_i,\bar x_i)&=&\sum_j\sigma^{pT}_{3,2}(x_i,\bar x_i)\big|_{j}\simeq\sum_j\sigma^{pD}_{3,2}(x_i,\bar x_i)\big|_{j}\frac{{\cal I}_T(X_j)}{{\cal I}_D(X_j)}
\end{eqnarray}

\noindent
and one may thus relate the 'known' quantity ${\cal R}_T'$ to the unknown properties of the hadron structure, represented by $K_{x_i}$ and $C_{3,j}(x_i,\bar x_i)$\ :

\begin{eqnarray}
{\cal R}_T'(x_i,\bar x_i)=
\frac{\sum_j\sigma^{pT}_{3,3}\big|_j(x_i,\bar x_i)}{\prod\sigma_S(x_i,\bar x_i)}\simeq K_{x_i}\sum_j{\cal N}'_jC_{3,j}(x_i,\bar x_i){\cal J}(Y_{j,i=1,3})
\label{eq:R'T}
\end{eqnarray}

Analogously to the case of double parton scattering, an indication on triple correlations in fractional momenta and in the transverse coordinates can then be obtained looking at the variation of ${\cal R}_T'$ as a function of $Y_{j,i}$ and using the property that $C_{3,0}=1$.

Notice that Eq.\eqref{eq:R'T}  is a consequence of Eq.\eqref{eq:S3,21}, which holds up to terms of ${\cal O}\Bigl((S_D^2-S_T^2)/S_T^6\Bigr)$. Since the right hand side of Eq.\eqref{eq:R'T} is of ${\cal O}\Bigl(1/S_T^4\Bigr)$, one may estimate that the relative correction to the dominant terms in Eq.\eqref{eq:R'T} is only of  ${\cal O}\Bigl((S_D^2-S_T^2)/S_T^2\approx1/5\Bigr)$. Eq.\eqref{eq:R'T} can therefore provide only semi-quantitative indication on the size of triple correlations, while a better determination requires a dedicated study.

\section{Final Discussion}

MPI in $pA$ collisions allow obtaining information on multi-parton correlations, which cannot be provided by studying MPI in $pp$ collisions\cite{Strikman:2001gz}.
Relevant features of the simplest case, double parton interactions in $pD$ collisions, were pointed out in \cite{Calucci:2010wg}. In the present paper, we have extended the study of MPI in $pA$ collisions to the cases of double and triple parton interactions in collisions of protons with $D$, $^3H$ and $^3He$, including in the discussion also the effects of interference terms.

Double parton interactions in collisions of protons with $D$, $^3H$ and $^3He$ are discussed in Sec.2. When only a single nucleon takes part to the hard process (Sub-Sec.2.1), the integrations on the relative transverse coordinates of the spectator nucleons are decoupled from all other transverse variables and the cross section is the same as measured in nucleon-nucleon collisions; apart from the proper multiplicity factor and the smearing corrections in the longitudinal variables, which, as discussed in Section 4, are however rather small. The explicit expression of the cross sections for $pD$, $p\ ^3H$  and $p\ ^3He$  collisions are given in Eq.10. Similar considerations hold in the case of triple parton collisions on a single nucleon, which is discussed in Sub-Sec.3.1. The corresponding contribution to the triple parton scattering cross section is given in Eq. 21. Notice that the spectator nucleons are on mass shell. As already discussed in \cite{Calucci:2010wg}, in spite of that, one may still claim that final state interactions of the spectators are taken approximately into account. The statement is supported by unitarity: If a nucleon is produced on mass shell and undergoes a final state interaction with the remnants of another nucleon, final state interaction does not modify the inclusive cross section, since the spectators are not observed. If a nucleon is produced off mass shell, its virtuality is anyhow rather small and it may not be unreasonable to extend the unitarity relation $SS^{\dagger}=1$ to the actual kinematical domain. Unitarity allows hence replacing all final state interaction with $cut$ nucleon lines, $i.e.$ with on mass shell nucleons.

In Sub Section 2.2, we discuss the case of double parton collisions, with two nucleons taking part to the hard process. In addition to the diagonal contribution in Fig.2a discussed in \cite{Calucci:2010wg}, which leads to the geometrical picture of the interaction in transverse space shown in Fig.3, one has a contribution from the non-diagonal contribution in Fig.2b. The geometrical picture in transverse space, of the corresponding interfering configurations a) and b), is shown in Fig.4. Notice that, in both interfering configurations, the hard interactions are localized in the same points and are well separated in transverse space. As a consequence, the argument for the suppression of the interference terms in MPI, discussed in \cite{Calucci:2009ea}, does not apply in this case. Differently from $pp$ collisions, the additional degrees of freedom provided by the nucleus, namely the possibility of having different nucleons involved in the hard process, can in fact produce the same partonic initial state in different ways, which can thus interfere in the process. As discussed in sub section 2.2, the contribution of the interference term is important in the region where the fractional momenta of the interacting partons are of order $\sqrt{E_B/m}$, where $E_B$ and $m$ are the nuclear binding energy and nucleon mass respectively. Differently from the diagonal contributions, which are dominated by the most probable nuclear configuration, where all nucleon's fractional momenta are equal, in the off-diagonal term of Fig.2b the fractional momenta $ Z_1$, $Z'_1$ and $Z_2$, $Z'_2$, of the two nucleons taking part to the hard process, are forced by kinematics to be different:  $ Z_1-Z'_1=-(Z_2-Z'_2)=\bar x_1-\bar x_2 $. Here $\bar x_1,\ \bar x_2$ are the fractional momenta of the two target partons undergoing the double collisions.
\par
As discussed in sub-section 4.1, the dominant contribution to the interference term can be expressed in terms of off diagonal parton distributions, Eq.s\eqref{eq:W1},\eqref{eq:H}. The contribution of the interference term can be singled out by looking at the dependence of the double parton scattering cross section on the difference $\bar x_1-\bar x_2$ (cfr. Eq.\eqref{eq:R1}). Keeping into account that the scale, which characterizes the dependence of the nuclear wave function on $Z$, is $\sqrt{E_B/m}$, one may roughly estimate that, to single out the contribution of the interference term one needs to measure the double parton scattering cross section in an interval $\bar x_1-\bar x_2\approx 5\times10^{-2}$ with an accuracy grater than 10\%. The whole discussion assumes that each couple of scattered partons, and the resulting observed particles, can be identified as a definite pair, which requires that each couple is sufficiently separated in phase space from the other couples. The quantitative amount of this separation depends on the detailed properties of the final state.

\par
As mentioned in sub-section 4.1, by studying the ratios in Eq.\eqref{eq:R} and in Eq.\eqref{eq:R1}, using the information on double parton interactions in $pp$ collisions and keeping into account that the dominant contribution to the diagonal term depends only on $K_{x_1,x_2}$, one may obtain information on parton correlations in fractional momenta and, through the overlap integrals in the transverse parton coordinates, which characterize $\sigma_{eff}$ and the interference term, also on parton correlations in the transverse coordinates. When the interference term is absent, as for production of $W$+ jets, the task of estimating parton correlations may be simpler. In that case, as discussed in\cite{Calucci:2010wg}, all information concerning correlations may be obtained directly from Eq.\eqref{eq:R}.
\par
The case of triple parton interactions has many features similar to the case of double parton interactions. The main difference is in the sizably larger number of contributing terms. As already pointed out, the contribution where only a single nucleon participates to the hard process is well approximated by the cross section on a isolated nucleon, multiplied by the multiplicity of target nucleons, while nuclear smearing effects can give only minor corrections. Things become complex when two or three nucleons participate to the hard process. The general features of the contributions where two nucleons participate to the hard process is discussed in Sub Section 3.2 while, in Sub-sub Section 4.2.1, one may find a simplified estimate of the different terms. In the case of two participating nucleons, one finds a diagonal and three different off diagonal contributions. Differently from the case of double parton collisions, in the case of triple parton collisions on two different target nucleons, the dominant contribution to the diagonal term depends both on the correlations in fractional momenta, through $K_{x_1,x_2,x_3}$, and on the correlations in the transverse coordinates, through the overlap function ${\cal G}_{x_i,\bar x_k}$ (defined by Eq.\eqref{eq:d32}). A consequence is that one cannot disentangle the effects of longitudinal and transverse correlations in triple parton collisions by studying  the ratios in Eq.\eqref{eq:R1D} and in Eq.\eqref{eq:R1D1} in $p\ D$ interactions only.
\par
The information on triple parton collisions in $p\ D$ interactions can be, nevertheless, utilized to estimate the contribution to triple parton collisions with two participating nucleons, in the case of $p\ ^3H$ or $p\ ^3He$ collisions. By measuring the triple parton scattering cross section in $p\ ^3H$ or $p\ ^3He$ collisions, one may thus estimate, using Eq.\eqref{eq:S3,21}, also the contribution to the cross section where three nucleons are involved in the hard process and thus figure out the value of the ratio ${\cal R}_T'$ in Eq.\eqref{eq:R'T}. A relevant feature of $\sigma^{pT}_{3,3}$, the component of the triple parton scattering cross section with three participating target nucleons, is that the leading contribution to the diagonal term, as given by Eq.\eqref{eq:S33}, is proportional to $K_{x_1,x_2,x_3}$ and does not depend on the correlations in the transverse coordinates. By studying the dependence of ${\cal R}_T'$ on $Y_{i,j}$ one can thus obtain an estimate both of $K_{x_1,x_2,x_3}$ and of the different overlap integrals in the transverse coordinates, which characterize the different interference terms. As discussed in the last part of sub-sub section 4.2.2, the uncertainties in the determination of $\sigma^{pT}_{3,2}$, by means of Eq.\eqref{eq:S3,21}, can however allow only a qualitative estimate of triple parton correlations, while a better determination requires a dedicated study. Some preliminary results of these investigations were presented at the Conference "New trends in high-energy physics - Alushta(Crimea) - 2011"\cite{Crimea}

\section{Conclusions}

The aim of the present paper is to study the possibility of obtaining model independent information on multi-parton correlations, by measuring MPI in high energy collisions of hadrons with light nuclei. Two different kinds of correlations, in fractional momenta and in the transverse coordinates, are in fact unavoidably linked and cannot be disentangled, when studying MPI in $pp$ collisions only. Already the simplest case, namely Double Parton Interactions in $pD$ collisions, is characterized by novel and non trivial features, as compared to DPI in $pp$ collisions. The component of the cross section, where both target nucleons contribute to the process, depends in fact only weakly on the correlations in transverse space. In addition, one has also a contribution from a interference term. All different contributions to the cross section can be disentangled and all the new unknown quantities appearing in the reaction and directly related to parton correlations, in fractional momenta and in the transverse coordinates, can be isolated in a way essentially model-independent. An interesting feature is that the interference term is expressed through the off diagonal parton distributions. By studying the interference term one may thus gain information also on the off diagonal parton distributions, in kinematical ranges not easily accessible through other processes.

\par
In order to disentangle triple parton correlations in fractional momenta and in the transverse coordinates, one needs to measure Triple Parton Interactions in $p\ ^3H$ or $p\ ^3He$ collisions. In fact, the knowledge of the cross section of Double and of Triple Parton Interactions in $pp$ and in $p\ D$ collisions is not sufficient to isolate all the unknown quantities, which appear in the reaction. On the other hand, keeping into account that the radii of $D$ and of $^3H$ and $^3He$ are not very different, Triple Parton Interactions in $p\ D$ collisions can teach a lot on the contribution to Triple Parton Interactions in $p\ ^3H$ or $p\ ^3He$ collisions, when only two target nucleons play an active role in the process. As discussed in Sub-Section 4.2, by taking advantage of the information on Triple Parton Interactions in $pp$ and $p\ D$ collisions, it's in fact rather simple to figure out how to obtain, from the cross section of Triple Parton Interactions in $p\ ^3H$ or $p\ ^3He$ collisions, a model-independent, although only qualitative, indication on the different components of the cross section, with a direct link either with the triple parton correlations in fractional momenta or with the triple parton correlations in the transverse coordinates.

\par
Our conclusion is that MPI of hadrons with light nuclei have a great potential to provide model independent information on the multi-parton structure of the hadron. By measuring the cross sections with a given number of MPI on various nuclear targets, one may in fact identify different features of the incoming parton flux, which allows isolating diverse terms of the correlated multi-parton structure. To our knowledge, a result not accomplishable by other means. The option of studying MPI in collisions of protons with light nuclei at RHIC and to run, at some stage, light nuclear beams at the LHC could thus be highly rewarding, offering the possibility to exploit the remarkable potential of MPI in $pA$ collisions to yield information on the many-body parton correlations and thus to provide unprecedented insight into the three dimensional structure of the hadron.

\vskip.5in

 \appendix
\section{\bf The non-relativistic three-body wave function}\par
 The nuclear systems we have considered (${}^3H,\;{}^3He$) can be treated with a
 non relativistic dynamics in their center-of-momentum frame, but since they are
 involved in a highly relativistic process it is necessary to match this
 internal non relativistic dynamics with the overall relativistic treatment.
 To this end the original procedure used by Salpeter\cite{Salpeter:1}\cite{Salpeter:2} to reduce the Bethe-Salpeter
 equation to the Schr\"odinger equation will be followed as strictly as possible. We are able only to set a correspondence between nonrelativistic and relativistic wave functions, but not to build up a wholly deductive procedure as in the quoted refernces. For simplicity we treat both the constituent and the bound state as a spinless boson. The final aim is to use the non relativistic wave functions, as they are known from nuclear physics, in our relativistic
 calculation with the correct factors and the correct kinematical transformation.

 \par
 The starting point is given by an homogeneous equations in relativistic form as obtained by a Feynman graph representation, in term of the two-body scattering matrices as suggested by the Faddeev\cite{Faddeev}\cite{Glockle} treatment of three-body ${\sf t_i}$ scattering:
\begin{eqnarray}
U_3(q_1,q_2,q_3)&=&\Delta(q_1)\Delta(q_2)\sum_{J\neq 3}\int i{\sf t_3}(q_1+q_2,k)U_J(q_1-k,q_2+k,q_3)dk,\cr
{\sf t_3}&=&V_3+G°_3 V_3{\sf t_3}\qquad G^o_3=\Delta(q_1)\Delta(q_2)
\end{eqnarray}
together with the two analogous terms for $U_1$ and $U_2$. An iteration of the above equation shows that all the three terms $U_J$ have in front the product of the free propagators of the constituent particles $\Delta(q_1)\Delta(q_2)\Delta(q_3)$, as explicit in the graphical description in Fig.10.

\begin{figure}[h]
\centering
\includegraphics[width=135mm]{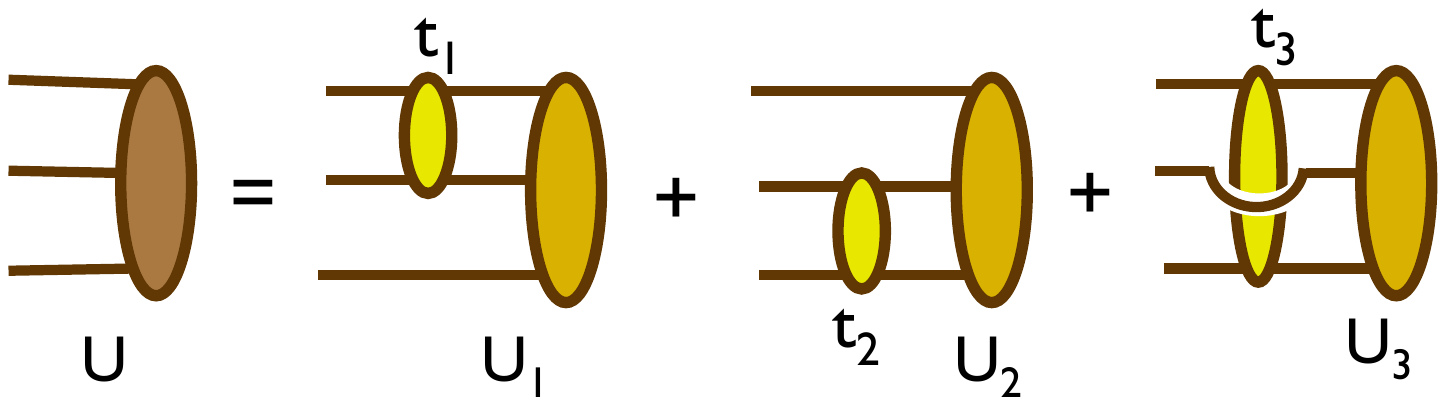}
\caption {Faddeev equation} \label{Faddeev}
\end{figure}
\par
 Defining $T=q_1+q_2+q_3$ the system is non-relativistic in the frame ${\mathbf T}=0$.\par
When the two-body scattering matrices $t_J$ do not depend on the relative energies but only oh the three-momenta, as it happens in nonrelativistic dynamics, we can integrate $U_J$ in $k_o$. A generalization of this procedure suggests the Ansatz:
\begin{eqnarray}
U_J(q_1,q_2,q_3)=i\Delta(q_1)\Delta(q_2)\Delta(q_3)\Phi_J(\mathbf {q_1,q_2,q_3})
\end{eqnarray}
This form has the meaning that, in the nonrelativistic limit, the bound particles are near the mass shell, so the singularities of the $\Delta-$factors are the most important; morover, in this limit, antiparticles are not
relevant, so the contribution of the antiparticle poles do not need to be taken into account in the integrations, explicitly:
$$\Delta(q)=\frac{i}{q^2-m^2}=\frac{i}{2\omega}\Big[\frac{1}{q_o-\omega}-\frac{1}{q_o+\omega}\Big]\approx \frac{i}{2\omega}\frac{1}{q_o-\omega}$$
By inserting eq.(A2) into eq.(A1), dropping the common factors $\Delta(q_1)\Delta(q_2)$ and integrating in $k_o$ we obtain:
\begin{eqnarray}
&\Delta(q_3)&\Phi_3(\mathbf{q_1,q_2,q_3})\cr
=&\Delta(q_3)&\sum_{J\neq 3}\int i {\sf t_3}(q_1+q_2,{\mathbf k})\Delta(q_1-k)\Delta(q_2+k) \Phi_J(\mathbf {q_1-k,q_2+k,q_3})dk\\
=&\Delta(q_3)& \sum_{J\neq 3}\int \frac{2\pi}{4\tilde \omega_1\tilde\omega_2}\frac{1}{(T-q_3)_o-\tilde \omega_1-\tilde\omega_2+i\epsilon}
 {\sf t_3}(q_1+q_2,{\mathbf k})
\Phi_J(\mathbf{ q_1-k,q_2+k,q_3})d^3k\nonumber
\end{eqnarray}
 The sub-energies are: $\omega_i=[(\mathbf {q_i})^2+m^2]^{1/2},\ \tilde\omega_1=[(\mathbf {q_1-k})^2+m^2]^{1/2},\;\tilde\omega_2=[(\mathbf{q_2+k})^2+m^2]^{1/2} $,
 We proceed by integrating over $(q_3)_o$.  Three sources of singularities need to be considered: \par\noindent
- singularities of the propagators where  $(q_3)_o$ appears directly; \par\noindent
- singularities of the propagators containing $ (q_1)_o, (q_2)_o $, where $(q_3)_o$ enters because we work at constant $T_o$. Actually it is convenient to make the position $q_1=(T-q_3)/2+l,q_2=(T-q_3)/2-l; $\par\noindent
- singularities of ${\sf t_3}$, possibly originating from the two-body subsystem (1+2), which can be either poles like $C/[(T-q_3)_o-\eta+i\epsilon]$ (two-body bound states) or cuts like
 $\int \rho(\eta)d\eta/[(T-q_3)_o-\eta+i\epsilon]$ (two-body scattering states).\par\noindent
A direct inspection shows that the pole $(q_3)_o=\omega_3-i\epsilon$ has imaginary part with sign opposite to all other poles in $(q_3)_o$, so its contribution gives the whole result of the integration. One obtains

 \begin{eqnarray}
&&\Phi_3(\mathbf {q_1,q_2,q_3})\\
&&=2\pi \sum_{J\neq 3}
\int d^3 k \frac{1}{4\tilde\omega_1\tilde\omega_2} {\sf t_3}(T-q_3,{\mathbf k})
\frac{1}{T_o-\tilde\omega_1-\tilde\omega_2-\omega_3+i\epsilon}
\Phi_J(\mathbf {q_1-k,q_2+k,q_3})\nonumber
\end{eqnarray}
 It is useful to make the positions:
 $T_o=M_T=3m+E_B,\;\omega_J=m+\kappa_J$, where by construction it results $E_B<0,\;\kappa_J>0$ . 
 For the function defined as:
 \begin{eqnarray}
\varphi_3(\mathbf {q_1,q_2,q_3})= \frac{\cal N} {E_B-\kappa_1-\kappa_2-\kappa_3}
\Phi_3(\mathbf {q_1,q_2,q_3}\;.)\end{eqnarray}
 one obtains the equation:
$$
\varphi_3=2\pi G_o(E_B)\sum_{J\neq 3}\int {\sf t_3} \varphi_J d^3 k \quad{\rm with}\quad  G_o(E_B)= \frac{1}{E_B-\kappa_1-\kappa_2-\kappa_3}\ ,
$$
\noindent
which, together with the two analogous equations for $\varphi_1,\varphi_2$, represent the usual system of Faddeev equation for a non-relativistic three-body system\cite{Faddeev}\cite{Glockle}, in fact $E_B$ is the binding energy and the $\kappa_J $ are the nonrelativistic kinetic energies.

\par
Defining as usual
$\mathbf{q_1=T/3+p_s/2+l,\;q_2=T/3+p_s/2-l,\;q_3=T/3-p_s}$,
 in the $\mathbf {T}=0$ frame, the wave function depends on two three-vectors and setting:
  $$\sum_J\varphi_J(\mathbf {q_1,q_2,q_3})=\varphi (\mathbf {p_s,l})$$
 $\varphi$ is the complete non-relativistic bound-state wave function, which can be made real and we must satisfy the normalization condition
 \begin{eqnarray}
  \int \varphi^2 d^3p_sd^3l=1\;.
 \end{eqnarray}

 The relations (A1,A4,A5) are linear and cannot give the normalization constant $\cal N$. A normalization condition may be obtained by considering the total charge of the bound state.

 \begin{figure}[h]
\centering
\includegraphics[width=135mm]{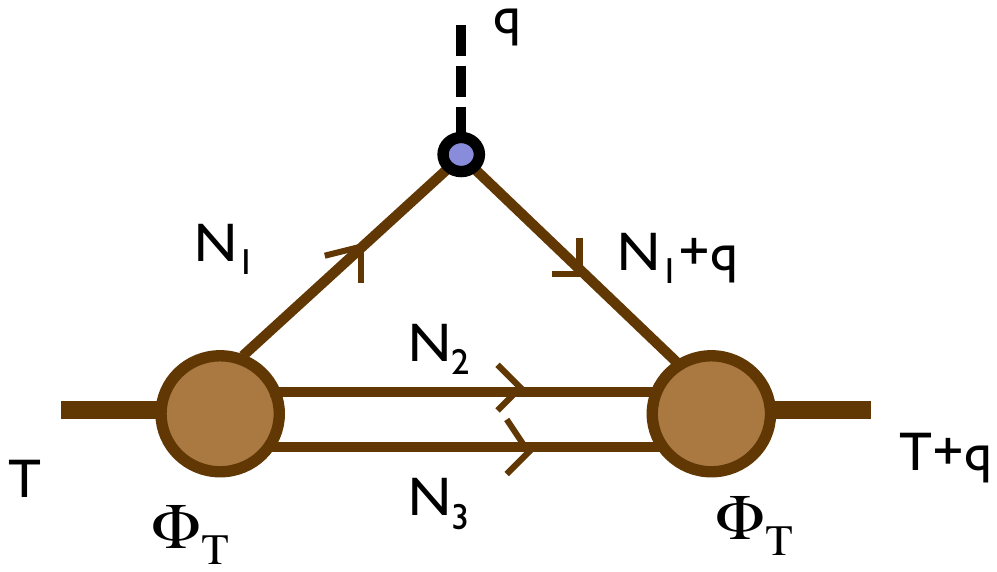}
\caption {Photon Tritium vertex} \label{Form factor}
\end{figure}

\par
  The coupling of the Tritium with an external electromagnetic field is written as
 \begin{eqnarray}
 {\cal J}^{\mu}(k)&=& 2T^{\mu}f(k^2)\cr &=&\int\Phi(\mathbf{q_1,q_2,q_3})\Delta(q_1)2q_1^{\mu}\Delta(q_1+k)\cr
 &&\qquad\qquad\qquad\times\Delta(q_2)\Delta(q_3)\Phi(\mathbf{q_1+k,q_2,q_3} )\delta (T-\sum q)\prod dq.
 \end{eqnarray}
 In the limit $k\to 0$ the zero component must give the total charge of the bound state
 $ {\cal J}_{0}(0)=2T_o$, or, in other words, the form factor must satisfy: $f(0)=1$.
 Now we use the relations $q_1=T-q_2-q_3$ and then, keeping the prescription of neglecting the antiparticle poles, we set:
 $$ \Delta(q)^2 \approx -\frac {1}{4q_o\omega}\Big[\frac {1} {q_o-\omega}\Big]^2$$
 The integration over the particle poles in ${q_2}_o\;,{q_3}_o $, is performed in the frame $\bf T$=0 with the result:
 \begin{equation}
 {\cal J}_{0}(0)=-(2\pi)^2\int \prod\frac {d^3q}{2\omega} \Phi(\mathbf{q_1,q_2,q_3})\bigg[\frac {1}{T_o-\omega_1-\omega_2-\omega_3}\bigg]^2 \Phi(\mathbf{q_1,q_2,q_3})\delta(\sum \mathbf {q})=2M_{T}
 \end{equation}
 We compare (A8) with (A6), at first sight we might conclude that we need an
$\cal N$ that depends on $\omega_i$, but we remember that in the present treatment we neglect $\kappa$ with respect to $m$, but not with respect to $E_B$, so we have finally
 \begin{equation}
{\cal N}=2m\sqrt{mM_{T}}/\pi
 \end{equation}
 The function $\varphi$ has dimensions -3 in powers of $q$, see (A6),
 then the function $\Phi$ has power $zero$ in $q$ as it must be, in fact $\Phi$ plays finally the role of an effective coupling in a four-boson relativistic vertex. Summing up the procedure: we start from $\varphi$ as it is known from nuclear physics, we construct then $\Phi$, eq.s (A5, A9) and from it we obtain, eq.(6), the expression of $\Psi$ that, after Fourier transformation, enters in the expression of the cross sections.\par
 The functions $\varphi$ are given in terms of the three dimensional momenta in the c.m. of the nucleus ${\bf q}_i$, we need them to be expressed in terms of the light cone fractional momenta to evaluate the cross sections in the main text. In the case of interest two nucleons are on shell (see the figure).

\begin{figure}[h]
\begin{center}
\includegraphics[width=120mm]{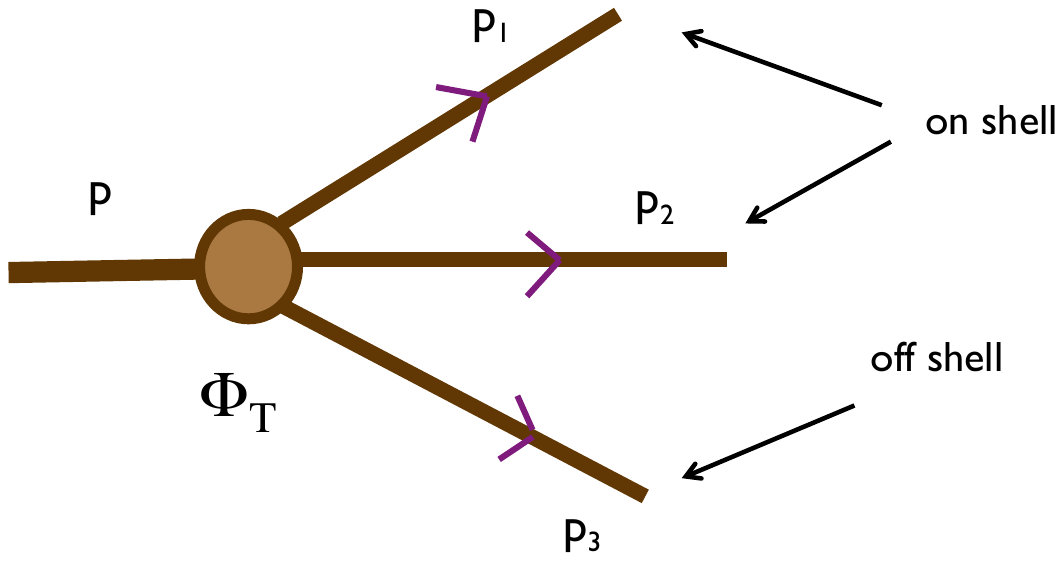}
\caption {$T/^3He$ vertex, one nucleon is virtual and two nucleons are on shell}
\label{Deuteron splitting}
\end{center}
\end{figure}

\noindent
Introducing the invariants

\begin{eqnarray}
s_{12}&=&(p_1+p_2)^2=(T-p_3)^2\nonumber\\
s_{23}&=&(p_2+p_3)^2=(T-p_1)^2\\
s_{13}&=&(p_1+p_3)^2=(T-p_2)^2\nonumber
\end{eqnarray}

\noindent
one has

\begin{eqnarray}
&&p_1^2=p_2^2=m^2,\qquad T^2=M_T^2,\qquad p_3^2\neq m^2\nonumber\\
&&s_{12}+s_{23}+s_{13}=M_T^2+2m^2+p_3^2
\end{eqnarray}

\noindent
The light-cone 4-momenta components are given by

\begin{eqnarray}
T&\equiv&\Biggl(\frac{M_T^2}{T_-},\ T_-,\ 0\Biggr)\nonumber\\
p_1&\equiv&\Biggl(\frac{m_{1\bot}^2}{Z_1(T_-/3)},\ Z_1T_-/3,\ {\bf p}_{1\bot}\Biggr)\nonumber\\
p_2&\equiv&\Biggl(\frac{m_{2\bot}^2}{Z_2(T_-/3)},\ Z_2T_-/3,\ {\bf p}_{2\bot}\Biggr)\\
p_3&\equiv&\Biggl(\frac{p_3^2+({\bf p}_{1\bot}+{\bf p}_{2\bot})^2}{(3-Z_1-Z_2)(T_-/3)},\ (3-Z_1-Z_2)T_-/3,\ -({\bf p}_{1\bot}+{\bf p}_{2\bot})\Biggr)\nonumber
\end{eqnarray}

\noindent
where $m_{i\bot}^2=m^2+{\bf p}_{i\bot}^2$ are the transverse masses.
The 4-momentum conservation

\begin{eqnarray}
M_T^2=\frac{3}{Z_1}m_{1\bot}^2+\frac{3}{Z_2}m_{2\bot}^2+\frac{3}{3-Z_1-Z_2}\bigl[p_3^2+({\bf p}_{1\bot}+{\bf p}_{2\bot})\bigr]^2
\end{eqnarray}

\noindent
implies

\begin{eqnarray}
p_3^2-m^2=(3-Z_1-Z_2)\Bigl[\frac{M_T^2}{3}-\frac{m_{1\bot}^2}{Z_1}-\frac{m_{2\bot}^2}{Z_2}-\frac{m_{3\bot}^2}{3-Z_1-Z_2}\Bigr]
\end{eqnarray}

\noindent
In the $T/^3He$ centre of mass frame, the nucleon's energies $E_i$, $i=1,2,3$ are expressed in terms of the invariants $t_i$ as follows

\begin{eqnarray}
E_{1}&=&\frac{M_T^2+m^2-s_{23}}{2M_T}\nonumber\\
E_{2}&=&\frac{M_T^2+m^2-s_{31}}{2M_T}\\
E_{3}&=&\frac{M_T^2+p_3^2-s_{12}}{2M_T}\nonumber
\end{eqnarray}

\noindent
The relations

\begin{eqnarray}
s_{23}&=&M_T^2+m^2-M_T^2\frac{Z_1}{3}-m_{1\bot}^2\frac{3}{Z_1}\nonumber\\
s_{31}&=&M_T^2+m^2-M_T^2\frac{Z_2}{3}-m_{2\bot}^2\frac{3}{Z_2}
\end{eqnarray}

\noindent
allow to express $E_{1}$ and  $E_{2}$ in terms of fractional momenta and transverse masses

\begin{eqnarray}
E_{1}&=&\frac{1}{2M_T}\Bigl(M_T^2\frac{Z_1}{3}+m_{1\bot}^2\frac{3}{Z_1}\Bigl)\nonumber\\
E_{2}&=&\frac{1}{2M_T}\Bigl(M_T^2\frac{Z_2}{3}+m_{2\bot}^2\frac{3}{Z_2}\Bigl)
\end{eqnarray}

\noindent
By using $E_i^2={\bf q}_i^2+m^2$ one obtains similar expressions for  centre of mass 3-momenta $q_{1z}$ and  $q_{2z}$

\begin{eqnarray}
q_{1z}&=&\frac{1}{2M_T}\Bigl(M_T^2\frac{Z_1}{3}-m_{1\bot}^2\frac{3}{Z_1}\Bigl)\nonumber\\
q_{2z}&=&\frac{1}{2M_T}\Bigl(M_T^2\frac{Z_2}{3}-m_{2\bot}^2\frac{3}{Z_2}\Bigl)
\end{eqnarray}

\noindent
Keeping into account

\begin{eqnarray}
M_T=E_1+E_2+E_3,\qquad{\bf q}_1+{\bf q}_2+{\bf q}_3=0
\end{eqnarray}

\noindent
one obtains the analogous relations for $E_3$ and $q_{3z}$:

\begin{eqnarray}
E_{3}&=&\frac{1}{2M_T}\Bigl[M_T^2\Bigl(2-\frac{Z_1}{3}-\frac{Z_2}{3}\Bigr)-3\frac{m_{1\bot}^2}{Z_1}-3\frac{m_{2\bot}^2}{Z_2}\Bigl]\nonumber\\
q_{3z}&=&\frac{-1}{2M_T}\Bigl[M_T^2\Bigl(\frac{Z_1}{3}+\frac{Z_2}{3}\Bigr)-3\frac{m_{1\bot}^2}{Z_1}-3\frac{m_{2\bot}^2}{Z_2}\Bigl]
\end{eqnarray}

\noindent
which allow to express explicitly the non-relativistic nuclear wave functions as a function of the fractional momenta $Z_i$ and of the transverse momenta ${\bf p}_{i\bot}$, i.e. in terms of variables invariant under longitudinal boost.

\section{ \bf Models worked out completely}

 We present here two models where more explicite calculations are carried out after having introduced more or less strong simplifications of the real dynamics.
\subsection{A model with the Hulth\'en wave-function}
 Since the Hulth\'en potential, with its relative wave function, is one of the simplest potential used in preliminary analyses of the Deuteron properties we present here a short derivation of some properties which are relevant for our investigations, in particular for an estimate of the relevance of the interference terms.\par
The relative two body wave-function for the ground state (S-wave), in $r$ and in $p$ representation is:
 \begin{equation}
 h({\bf r})=\frac{\sqrt{\kappa\tau
 (\tau+\kappa)/2\pi}}{\tau-\kappa}\frac{1}{r}\big[e^{-\kappa r}-e^{-\tau r}\big]
 \qquad
\tilde h({\bf p})=\frac{\sqrt{\kappa\tau (\tau+\kappa)}}{\pi(\tau-\kappa)} \bigg[\frac{1}{{\bf p}^2+\kappa^2}-\frac{1}{{\bf p}^2+\tau^2}\bigg]
\end{equation}
 where $\kappa=\sqrt{mE}\quad \tau=\kappa+\mu\quad$the binding energy is $E=2m-M_D$, $1/\mu$ should represent the range of the potential, really it has been fitted phenomenologically to $\mu\approx 5\kappa$ , $m/2$ is the reduced mass, assuming equal masses for the nucleons. The normalization is:
 $$4\pi\int |h({\bf r})|^2 r^2dr=1\qquad 4\pi\int |\tilde h({\bf p})|^2p^2dp=1$$

 Now we may calculate the mean value and the dispersion of the radial coordinate
 with the result:
  \begin{equation}
   <r>=\frac {\tau^2+4\tau\kappa+\kappa}{2\kappa\tau(\kappa+\tau)}\qquad
  <r^2>-<r>^2=
\frac{\tau^4+2\tau^3\kappa+6\tau^2\kappa^2+2\tau\kappa^3+\kappa^4}
{[2\tau\kappa(\kappa+\tau)]^2}
 \end{equation}
 In the actual case the parameters satisfy the condition $\kappa<<\tau$ and
 there is a strong simplification:
 $$<r>\approx\sqrt{<r^2>-<r>^2}\approx1/2\kappa$$
 The longitudinal variable $Z$ is studied in an analogous way
 and the results are:
 \begin{equation} <Z^2>=\frac{4}{M_D^2}\Big[\frac{4}{3}<{\bf p}^2>+m^2\Big]
 \qquad
  <Z^4>=\frac{16}{M_D^4}\Big[\frac{16}{5}<{\bf p}^4>+4m^2<{\bf p}^2>+m^4\Big]
  \end{equation}
 with the mean values:
 $$<{\bf p}^2>=\kappa\tau \quad <{\bf p}^4>=\kappa\tau (\kappa^2+3\kappa\tau+\tau^2)$$
 The qualitative behaviors of the parameters $B,\;Z$ we learn are that the transverse extension is as large as one expected and the relative dispersion is quite large; the longitudinal variable $Z$ is centered aroun 1, with a relatively small dispersion at least for $\kappa<\tau<m$ where we find:
 $$\sqrt{<Z^4>-<Z^2>^2}/<Z^2>=\sqrt{3\kappa\tau}/6m$$
 The fact the the fractional momentum is slightly larger then 1 is due to the
 fsact the we are, really, considering an unsymmetrical situation where one of
 the bound nucleon is put on mass shell, the configuration is symmetrical in the
 space variables $\bf p$, but it is not symmetrical in the relative energies.
 It seems that the more interesting result is the dispersion in $Z$, in fact
 this is the parameter which says how much the contribution of the interference
 term, where $Z\neq Z'$, differs from the diagonal term.

 We are also interested in a mixed representation where the transverse degrees of freedom are expressed in space variables $B$ while the longitudinal degree is given in light-cone variables $p_+,p_-$. We recall that we are interested in a particular kinematical situation where one of the bound nucleon is treated as real on mass shell but we still are on the center-of-momentum frame so that the two three momenta are opposite, then we shall consider a longitudinal boost. In this situation it results
 $p_z=m_{\bot}^2/2p_- -p_-/2\quad m_{\bot}^2=m^2+p_{\bot}^2$
 With these definitions we obtain:
 $$\frac{1}{{\bf p}^2+\alpha^2} =\frac{p_-}{\sqrt{m^2-\alpha^2}}\bigg[\frac{1}{p_{\bot}^2+w^2}-\frac{1}{p_{\bot}^2+v^2}\bigg]$$
 where
 $$v_{\alpha}^2=p_-^2+2p_-\sqrt{m^2-\alpha^2}+m^2\quad
 w_{\alpha}^2=p_-^2-2p_-\sqrt{m^2-\alpha^2}+m^2$$
 and $\alpha$ is either $\kappa$ or $\tau$.
 The definition $Z=2p_-/D_-$ gives in the center-of-momentum frame $p_-=M_DZ/2$
 and so we get:
\begin{eqnarray}
 \hat h(B,Z)= \frac{M_D}{\mu}\sqrt{\kappa\tau(\kappa+\tau)}&\bigg[&\frac{Z}{\sqrt{m^2-\kappa^2}}[K_o(w_{\kappa}B)-K_o(v_{\kappa}B)]\cr
&-&\frac{Z}{\sqrt{m^2-\tau^2}}[K_o(w_{\tau}B)-K_o(v_{\tau}B)]\bigg]\;.
 \end{eqnarray}

 In order to apply these expressions to the Deuteron's case we take into account that $\kappa<\mu<m$; we have already noted that we are interested in small values of $B$ compared
 with the nuclear scale $1/2\kappa$, so we look for the limit $B\to 0$ which gives $K_o(w_{\kappa}B)-K_o(v_{\kappa}B)\to \ln (v_{\kappa}/w_{\kappa})$.

\begin{figure}[h]
\centering
\includegraphics[width=110mm]{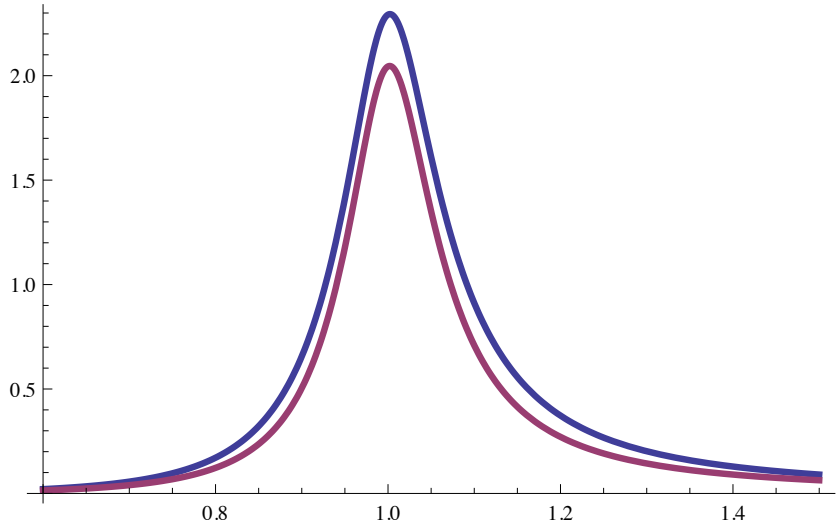}
\caption {Graph of $\hat h(B=0)$ as function of $p_-/m$.
Arbitrary vertical normalization; the upper curve corresponds to $\mu=3\kappa$, the lower curve corresponds to $\mu=5\kappa$.}
\end{figure}

 We can study numerically the above limiting form of $\hat h$, the result for
$E/m=0.0021$ are presented in the graph as function of $u=p_-/m$, which is slightly different from $Z=2mu/M_D $.
Since $\mu=5\kappa$ is a phenomenological fit without direct interpretation the numerical study has been performed also for $\mu=3\kappa$ which would better describe a binding potential generated by pion exchange; the qualitative conclusions are however the same with both choices of the parameters. In Fig.13 we plot $\hat h(0,Z)$ as a function of  $Z$ in the cases $\mu=3\kappa$ and
$\mu=5\kappa$. A numerical study shows that the shape of  $\hat h(B,Z)$ has little variation with increasing $B$.
In accordance with the previous results the spread in $Z$ is sizable, in order to go from the maximum  of $\hat h$ to half of this maximum $Z$ must vary at least by 0.1. So the product of two function with $Z$ different of some per cent, and then its integral, is not very depressed with respect to a diagonal term. This in practice means that the interference terms are smaller than the diagonal ones, but not by orders of magnitude.

 \subsection { A Gaussian model for the transverse dynamics}

 \subsubsection{General features of the model}
Here the simplification is heavier, in the aim of dealing explicitly with the transverse degrees of freedom for all the graphs we considered. No correlations among transverse and longitudinal degrees of freedom are taken into account, moreover the transverse distributions are Gaussian. One of the results, which seems us of more general validity is that the transverse dynamics is, in comparison with the longitudinal, less sensible to the difference between diagonal terms and interference terms.

\par
 The one-parton inclusive distribution is written as:

 \begin{equation}
 \Gamma_1=G(x)f_1(b)\quad f_1(b)=\frac{1}{\pi R^2}\exp[-b^2/R^2]\;.
\end{equation}

 The two-parton distribution is:

\begin{eqnarray}
\Gamma_2=K G(x_1) G(x_2) f_2(b_1,b_2) \quad f_2(b_1,b_2)&=&
\frac{1}{(\pi R^2)^2(1-\lambda^2)}
\exp[-(b_1^2+b_2^2+2\lambda b_1b_2)/R^2(1-\lambda^2)]\cr
\int f_2(b_1,b_2)db_2&=&f_1(b_1)
\end{eqnarray}

 $ K$ controls the parton multiplicity, $\lambda $ their spatial correlation; both can still depend on the fractional momenta $x_i$.

\par
The three-parton distribution, with the minimal number of new parameters, is written as:

\begin{eqnarray}
 \Gamma_3&=&G(x_1) G(x_2) G(x_3)K_3 f_3(b_1,b_2,b_3) \cr
  f_3(b_1,b_2,b_3)&=&\frac{1}{(\pi R^2)^3(1-2\lambda)(1+\lambda)^2} \exp\Big[-\frac{(1-\lambda)\sum_i b_i^2+2\lambda\sum_{i<j}(b_i\cdot b_j)}{R^2(1+\lambda)(1-2\lambda)}\Big]
\end{eqnarray}

where we have implemented the requirement that $f_3$ be symmetric in the parton variables and that
 $$\int f_3(b_1,b_2,b_3)db_3=f_2(b_1,b_2)\;.$$
\par
 The information embodied in the parameters $K_N$ and $\lambda$ can be seen in
 this way. The distributions $f_N$ are inclusive, if the corresponding exclusive distributions, integrated on the longitudinal variables in a given interval, were Poissonian, then
 $K_N$ would be 1. Of course even a
 Poissonian distribution of the partons' multiplicities can imply spatial  corralations if $\lambda\neq 0$.

 \par
 The nuclear distribution, $i.e.$ the square of the wave function are expressed for the deuteron as:

\begin{equation}
 g(z;b)=W(Z)F(B)\quad  F(B)=\frac{1}{\pi S^2}\exp[-B^2/S^2]\qquad 0<Z<2\quad B=B_1-B_2
\end{equation}

 Since the distribution will be at the end integrated, for the Tritium we write:

 \begin{eqnarray}
 g(Z_i,B_i)&=&W(Z_1,Z_2,Z_3)\delta(Z_1+Z_2+Z_3)F(B_i)\qquad 0<Z<3 \cr
 F(B_i)&=&\frac{4}{3(\pi S^2)^2}\exp\big[-2[(B_1-B_2)^2+(B_2-B_3)^2+(B_2-B_3)^2]/3S^2\big]
 \end{eqnarray}

 The normalization is
 $$\int W(Z)dZ=1,\quad\int W(Z_1,Z_2,Z_3) \delta(Z_1+Z_2+Z_3)dZ_1dZ_2dZ_3=1 $$
 In nonrelativistic case $Z\approx 1$. The transverse three-body distribution satifies
$$\int F(B_i) dB_3=\frac {1}{\pi S^2}\exp[-B^2/S^2]\quad B=B_1-B_2 \;,$$
which defines the normalization, it has been already
 noted that the sizes $S$ in the deuteron and in the Tritium, although similar, are in fact different.
  \subsubsection{Free nucleons}

 As reference quantities we consider, within the model, the simple double and triple hard scattering among free nucleons, at fixed fractional momenta.
 The simple hard scattering is described by
 $$ \sigma_1(x,x')=\hat \sigma_{xx'}G(x)G(x')\int f_1(b)f_1(b-\beta)dbd\beta=
 \hat \sigma_{xx'}G(x)G(x')$$
 due to the normalization of the transverse distributions.
The double hard scattering is described by

 \begin{eqnarray}
\sigma_2(x_1,x_2;x'_1,x'_2)&=&K_2^2 \hat \sigma_{x_1x_1'}G(x_1)G(x'_1)
\hat \sigma_{x_2x_2'}G(x_2)G(x'_2) I_o\cr
I_o&=&\int f_2(b_1,b_2)f_2(b_1-\beta,b_2-\beta)db_1 db_2 d\beta\end{eqnarray}

The triple hard scattering is described by
 \begin{eqnarray}
&\sigma_3&(x_1,x_2,x_3;x'_1,x'_2,x'_3)=K_3^2
\hat \sigma_{x_1x_1'}G(x_1)G(x'_1)
\hat \sigma_{x_2x_2'}G(x_2)G(x'_2)
\hat \sigma_{x_3x_3'}G(x_3)G(x'_3)J_o\cr
J_o&=&\int f_3(b_1,b_2,b_3)f_3(b_1-\beta,b_2-\beta,b_3-\beta)db_1 db_2 db_3 d\beta
\end{eqnarray}
   All the transverse variables "$b$" are two-dimensional and the cylindrical symmetry is always preserved, so all the calculations regarding the transverse variables have this standard form
 \begin{equation}
 C\Big[\int d^N y\; \exp [-y\cdot {\sf M}\cdot y]\Big]^2=C\frac {\pi^N}{{\sf det\,M}}
\end{equation}
 Here ${\sf M}$ is a $N\times N$ matrix, $C$ embodies the normalizing factors as given in Eq.s (B5, B6, B7) \par
 In the cases considered above,
 after a rescaling of the variables, $b\to bR\sqrt{1-\lambda^2}$ in the first
 case and $b\to bR\sqrt{1+\lambda}\sqrt{1-2\lambda}$ in the scecond case the matrices take the form
\par
 \begin{eqnarray}
 \sf M_I=\left\|\begin{matrix}
 2& 2\lambda&  -1-\lambda &\hfill\cr
  2\lambda& 2 & -1-\lambda& \hfill\cr
                           -1-\lambda& -1-\lambda &2+2\lambda&\hfill\cr
                           \end{matrix}\right\|
                           \end{eqnarray}

 \begin{eqnarray}
 \sf M_J=\left\|\begin{matrix}
 2(1-\lambda)&  2\lambda&2\lambda& -1 & \hfill\cr
  2\lambda  & 2(1-\lambda) & 2\lambda &-1& \hfill\cr
                         2\lambda  &2\lambda&2(1-\lambda)&-1&\hfill\cr
                           -1 & -1 &-1&3(1+\lambda)&\hfill\cr
                           \end{matrix}\right\|
                           \end{eqnarray}
 So that we have, at the end:      \begin{eqnarray}
 I_o &=&\frac{1}{4\pi R^2} \frac{1}{1+\lambda} \cr
 J_o&=&\frac{1}{12(\pi R^2)^2} \frac{1}{1+4\lambda+2\lambda^2}
  \end{eqnarray}
 \subsubsection {Bound nucleons}

We start considering the scattering process when one of the nucleon is bound in a deuteron
$$ \sigma_{D,1}(x,x')=\hat \sigma_{xx'}G(x)G(x'/Z)W(Z)dZdxdx'\int  f_1(b)f_1(b-B_1) F(B_1-B_2)db dB_i$$
 By integrating in $B_2$ the factor $F$ one gets simply 1. According with the discussion in Section 3, in the distribution $W$ the variable $Z$ is shrunk around the value $Z=1$ so that we take  $\int G(x'/Z)dZ W(Z)\approx  G(x')\int dZ W(Z)= G(x')$ and we obtain the same expression as for the free case.
   From the simple procedure described is is seen that the same result holds for the case of the Tritium, with a suitable redefinition of the size $S$ also when we consider double or triple hard scatterings.\par
When we look at hard scatterings where more than one bound nucleon participates new features appear. In the simplified model here described the longitudinal degees of freedom are factorized we have anyhow seen that $Z\approx 1$ approximation is
inconsistent with the conservation of longitudinal momentum in the nondiagonal cases, but we do not have anything new to add to this point: we investigate the
transverse degrees of freedom of the partons (inside the nucleon) and of the nucleon (inside the nucleus) which are dynamically connected.
\par
Now we study the double scattering when both nucleons of the deuteron are involved:
(here and below the $x,x'$ arguments of $\hat\sigma$ will be usually omitted)\par
It has been shown that in this case there are two possibilities: the direct term and an interference term. In this last case we cannot work simply with the density distribution of partons but we need the "wave function", whose absolute square gives the distribution of the partons inside the hadron. In our case, where the distributions are Gaussian we take as "wave function" simply the square root of the distribution, $i.e.$ we ignore the possibile phases. Then the relevant quantities are:\par
 \begin{eqnarray}
 I_2^d&=&\int f_2(b_1,b_2)f_1(b_1-B_1)f_1(b_2-B_2)F(B_1-B_2)dbdB\cr
 I_2^i&=&\int f_2(b_1,b_2)\sqrt{f_1(b_1-B_1)}\sqrt{f_1(b_1-B_2)}\sqrt{f_1(b_2-B_2)}\sqrt{f_1(b_2-B_1)}F(B_1-B_2)dbdB\qquad
\end{eqnarray}
The subsequent calculations are very similar to the previous one. There is only a new feature, the new dimensional parameter $S$. We have an integration over the transverse variables $b$ and we define adimensional variables through $b\to bR$ and, for convenience we introduce $v=R^2/S^2$. The Gaussian integration involves the calculation of $4\times 4$ determinants and finally, taking into account the normalizations we get, for the diagonal and for the interference terms, respectively:\begin{equation}
I_2^d=\frac{1}{\pi[S^2+2(2+\lambda)R^2]} \qquad
I_2^i=\frac{1}{\pi(2+\lambda)[S^2+2R^2]}
\end{equation}
 In the double collision involving a  Tritium there is necessarily
     a spectator. Since by integrating the three-body distribution (B5) over the spectator's coordinates we obtain the two-body distribution (B4) the final expression is the same, provided we rescale, in (B11) the size according to the experimental values.
      \par
     Now we consider that triple hard interaction where one of the bound nucleons interacts twice, another only once and there is, in the Tritium case, a spectator.
   It has already been shown that there are two kind of processes in this case, a diagonal term and some interference terms which must be treated separately.\par
   We do not repeat the consideration about the longitudinal variables, for what concerns the transverse variables the integration implies the calculation of $5\times 5$ determinants and the final result is, for the diagonal term:
 \begin{equation}
   J_2^d=\frac {1}{1+\lambda}\frac {1}{4\pi^2 R^2[S^2+(3+\lambda)R^2]} \end{equation}
   For the interference terms we get
   	\begin{eqnarray}
  J_2^{i,2}&=&\frac {1-\lambda}{(1+\lambda)(6-\lambda)}\frac {4}{\pi^2 R^2  [(4-\lambda)S^2+2(4-2\lambda-\lambda^2)R^2]} \cr
  J_2^{i,3}&=&\frac {1-\lambda}{(1+\lambda)(3+\lambda)}\frac {1}{\pi^2R^2 [(3-\lambda)S^2+4(1-\lambda)R^2]} \cr
    J_2^{i,4}&=&\frac {1-\lambda}{(1+\lambda)(4-2\lambda-\lambda^2)}\frac {4}{\pi^2R^2 [7S^2+2(6-\lambda)R^2]}
     \end{eqnarray}
  Finally we consider the process where we have three bound nucleons, all interacting once, evidently now we must consider a Tritium. We find a diagonal term and two different interference terms, the integration implies the calculation of $6\times 6$ detrminants and the final result is, for the diagonal term:
    \begin{equation}
  J_3^d=\frac {4}{3}\frac {1}{\pi^2 [S^2+2(2+\lambda)R^2]^2} \end{equation}
  For the two interference terms we get:
  \begin{eqnarray}
  J_3^{i,2}&=&\frac {4}{3}\frac {1}{\pi^2 (2+\lambda)[(S^2+2R^2)(S^2+2(2+\lambda)R^2]}\cr
  J_3^{i,3}&=&\frac {4}{3}\frac {16}{\pi^2[(7+3\lambda)S^2+8(2+\lambda)R^2]^2}
   \end{eqnarray}
 \par
 A feature that appears very clearly from the model but that reflects a more general property is the dependence on the geometrical parameters: since the cross section has dimension $\ell^2$ and it embodies three factors $\hat\sigma$, with dimension $\ell^2$ each one there must be a factor $\ell^{-4}$. When only one nucleon interacts this factor is necessarily $1/R^4$, when two nucleon interacts the factor is $1/R^2(\mu S^2+\nu R^2)$, when three nucleons interact the factor has the form $1/(\mu S^2+\nu R^2)^2$. The meaning appear clear considering the hypothetical situation $S>>R$. In this case the first cross section remains unaltered, the second vanishes as $1/S^2$, the third one vanishes as $1/S^4$.

 \section{\bf Infrared behavior}
We give a short look at the infrared properties of the amplitudes and densities we have used.\par
Using eq.s (5,9) we would conclude that the one particle density has the behavior $\Gamma(z;b)\propto 1/z$, but we know that this is not supported by experimental data, so we must conclude that the vertex $\phi$ must show also an infrared singularity in such a way that $\Gamma(z;b)\propto 1/z^{1+\nu}$. Since the integration (with an infrared cutoff) of the two-body distribution must give the one-body distribution the same behavior must be found in the two-body vertex $\hat\phi$.\par
Thus we find a relevant simplification of the two-parton amplitude when one of the two partons has its four momentum very soft, it is the particular case of the general features found in the emission of soft particles. For definiteness we consider the free proton case: when the parton four-momentum goes to zero linearly in all its components, we have ${l_1}_{\bot}\propto x_1$. Then in the expression for $\psi_2$ we neglect the term in ${l_1}_{\bot}^2$ and also $x_1$ with respect to $x_2$ with the result.
 $$\psi_{i.r.} =\frac {1}{\sqrt 2 L_+x_1}\;
\frac {\hat\phi}{x_2[m^2-M_{\bot}^2/(1-x_2)]-{l_2}_{\bot}^2}$$
but we have seen that $\hat\phi$ must have a singularity, we extract it by writing $\hat\phi\propto \frac{\phi}{x^{\nu/2}}$.
 In the same kinematical configurations also the other factor used in defining the densities $\Gamma$ can be decomposed as
$$\frac {x_1,x_2}{1-x_1-x_2}\simeq x_1\frac {x_2}{1-x_2} \qquad x_1<<x_2 \;. $$
 So the original density is decomposed into two factors;  the second factor is precisely the probabiliy of finding one parton with finite fractional momentum, as seen in eq.(5),the first one can be thought as the usual infrared term of QED correced phenomenologically enhancing the singularity; the well known term of QED is $P\cdot\varepsilon/P\cdot q$, it could represent also a gluon emission, but since we have not taken into account the spin we do not have the numerator $P\cdot\varepsilon$; the denominator is dominated by the "large" component $x_1L_+$.\par
We use the mixed representation of $\psi$ with longitudinal momenta and transverse coordinate, so we should perform the Fourier transformation in $l_{\bot}$. The operation on ${l_2}_{\bot}$ yields precisely $\psi(x_2;b_2)$, on   ${l_1}_{\bot}$ the integration must run only on the infrared domain of ${l_1}_{\bot}$, limited by a cut off $|{l_1}_{\bot}|<\ell_{i.r.}$ for dimensional reasons
the result is proportional to a function $\ell_{i.r.}^2G(b_1\ell_{i.r.})$ and we get for the  density $\Gamma(x_1,x_2;b_1,b_2)\simeq\ell_{i.r.}^2 G(b_1\ell_{i.r.})\Gamma(x_2;b_2)/{2x_1}$. Integrating over the longitudinal infrared momentum $x_1$ we obtain the usual phenomenological divergence $dx/x^{1+\nu}$.\par
It is in fact a very general property that the soft emission is relatively independent by the rest of the dynamics, so it holds also for the more complicated expressions, like the terms $W_j$ and also for three-body amplitudes, which become completely factorized in the limit $x_1<<x_3;\;x_2<<x_3$. In detail the treatment applies directly to a soft emission originating directly from a free nucleon. When the parton emission comes from a nucleus, one finds the usual complication due to the binding, but when the treatment is applied to a configuration with $z<<1$, the previous discussion still holds, see eq.21.
 The factorization which is produced in this particular case can be useful in order to make a simpler relation, at least in this kinematical configuration, between the diagonal and the interference term.

\end{document}